 \documentclass[preprint]{aastex}

\shorttitle{Speckle Measures with DSSI. VII.}
\shortauthors{Horch et al.}

\begin{document}
\newcommand{\beq}{\begin{equation}}
\newcommand{\eeq}{\end{equation}}
\newcommand{\ea}{{\it et al.\ }}


\title{Observations of Binary Stars with the Differential Speckle
Survey Instrument. VII. Measures from 2010 September to 2012 February at the 
WIYN Telescope}

\author{Elliott P. Horch\altaffilmark{1,9,12}, Dana I. Casetti-Dinescu\altaffilmark{1},
Matthew A. Camarata\altaffilmark{1,10,12}, Akbar Bidarian\altaffilmark{1,12},
William F. van Altena\altaffilmark{2,12},
William H. Sherry\altaffilmark{3,11},
Mark E. Everett\altaffilmark{3},
Steve B. Howell\altaffilmark{4,12},
David R. Ciardi\altaffilmark{5},
Todd J. Henry\altaffilmark{6}, Daniel A. Nusdeo\altaffilmark{7}, and
Jennifer G. Winters\altaffilmark{8}}

\affil{$^{1}$Department of Physics,
Southern Connecticut State University,
501 Crescent Street, New Haven, CT 06515}

\affil{$^{2}$Department of Astronomy, Yale University,
P.O. Box 208101, New Haven, CT 06520-8101}

\affil{$^{3}$National Optical Astronomy Observatory,
950 North Cherry Avenue, Tucson, AZ 85719}

\affil{$^{4}$NASA Ames Research Center, Moffett Field, CA 94035}

\affil{$^{5}$NASA Exoplanet Science Institute, California Institute
of Technology, 770 South Wilson Avenue, Mail Code 100-22, Pasadena, CA 91125}

\affil{$^6$RECONS Institute, Chambersburg, PA, 17201}

\affil{$^{7}$Department of Physics and Astronomy, Georgia State University, 25 Park Place,
Atlanta, GA 30302}

\affil{$^{8}$Harvard-Smithsonian Center for Astrophysics, 60 Garden Street, Cambridge, MA 02138}

\email{horche2@southernct.edu, danacasetti@gmail.com, camarata@hhne.org,
bidariana1@southernct.edu, 
william.vanaltena@yale.edu, william.sherry@paradisevalley.edu,
everett@noao.edu,
steve.b.howell@nasa.gov, ciardi@ipac.caltech.edu,
thenry@astro.gsu.edu, nusdeo@astro.gsu.edu,
jennifer.winters@cfa.harvard.edu}

\altaffiltext{9}{Adjunct Astronomer, Lowell Observatory, 1600 West Mars Hill Road, 
Flagstaff, AZ 86001.}

\altaffiltext{10}{Current Address: Hebrew High School of New England, 
30 Bloomfield Avenue, West Hartford, CT 06117.}

\altaffiltext{11}{Current Addresses: Physical Sciences Department, 
Glendale Community College,
6000 West Olive Avenue, Glendale, AZ 85302 and
Eureka Scientific, Inc., 2452 Delmer Street, Suite 100,
Oakland, CA 94602-3017}

\altaffiltext{12}{Visiting Astronomer, Kitt Peak National Observatory, 
National Optical Astronomy Observatory, which is operated by the Association of 
Universities for Research in Astronomy, Inc. (AURA) under cooperative agreement 
with the National Science Foundation.}


\begin{abstract}

We report on speckle observations of binary stars carried out at the WIYN 
Telescope over the period from September 2010 through 
February 2012, providing 
relative astrometry for 2521 observations of 883 objects, 856 of which
are double stars and 27 of which are triples. The separations measured
span a range of 0.01 to 1.75 arc seconds.
Wavelengths of 562 nm, 692 nm, 
and 880 nm were used, and differential photometry at one or more of these
wavelengths
is presented in most cases. Sixty-six components were resolved for
the first time. We also estimate detection limits at 0.2 and 1.0 
arc seconds for high-quality
observations in cases
where no companion was seen, a total of 176
additional objects. Detection limits vary based on observing conditions and
signal-to-noise ratio, but are approximately 4 magnitudes at
0.2 arc seconds and 6 magnitudes at 1.0 arc seconds on average.
Analyzing the measurement precision of the data set, we
find that the individual separations obtained have 
linear measurement uncertainties of approximately 2 mas,
and photometry is uncertain to approximately 0.1 magnitudes in general.
This work provides fundamental, well-calibrated data for future orbit and
mass determinations, and we present three first orbits and total mass estimates
of nearby K-dwarf
systems as examples of this potential.

\end{abstract}

\keywords{binaries: visual --- 
techniques: interferometric --- techniques: high angular resolution ---
techniques: photometric}

\section{Introduction}

Binary stars remain an important tool in furthering two fundamental areas
of astrophysics: (1) they contribute to an understanding of stellar 
structure and 
evolution by providing empirically-determined masses of stars, and (2) 
when used statistically,
observationally-determined separations,
magnitude differences, and orbital parameters
can yield information important for
understanding the relative importance of star (and by extension, exoplanet)
formation mechanisms. Speckle imaging
has contributed in both areas over its long 
history, and indeed the technique has
seen a resurgence in recent years 
due to the more widespread use of electron-multiplying CCDs (EMCCDs)
in speckle work (Tokovinin and Cantarutti 2008; Horch \ea 2011a;
Docobo \ea 2010;
Balega \ea 2013). In addition to opening new parameter space for 
the measurement of binary systems due to their sensitivity and readout
speed, these devices have enabled  
the use of speckle imaging for faint (stellar) companion detection for stars 
thought to host exoplanets (Howell \ea 2011; Howell \ea 2016; Furlan \ea in press;
Hirsch \ea in press).
Speckle imaging has been used together with adaptive optics to
determine if a stellar companion is present, and if so, what its
brightness relative to the primary star is ({\it e.g.\ }Crossfield \ea 2016), 
which is an important consideration in deriving
the correct radius of the exoplanet (Ciardi \ea 2015).
Adaptive optics and speckle imaging are essentially complementary in this
regard because the bulk of adaptive optics observations have relatively 
shallow detection limits within 0.2 arc seconds due to the fact that 
the wavelengths of observation are usually in the infrared; in contrast,
speckle imaging provides diffraction-limited imaging over a very small 
field of view, but in visible wavelengths. Therefore, the detection limits
close to the target star are often deeper for a comparable telescope
aperture, something that enables astrophysics with close binary stars as well
as exoplanet host star vetting.

Our past work at the WIYN\footnotemark \hspace{4pt}telescope 
has involved both surveys of binary stars and exoplanet host
\footnotetext{At the time of these observations, the WIYN Observatory was 
a joint facility of the University of Wisconsin-Madison, Indiana
University, Yale University, and the National
Optical Astronomy Observatories.}stars.
In particular,
earlier papers in this series have detailed our complete survey of
{\it Hipparcos} double stars and suspected doubles, which began in the late
1990's (Horch \ea 1999), was active over the next fifteen years 
({\it e.g.\ }Horch \ea 2012, Horch \ea
2015a, and references therein), 
and has continued through the present work. 
This program has contributed a large amount of 
relative astrometry and photometry data as well as 
about two dozen orbits and total
mass estimates. Some of the smallest separation systems that we have successfully 
observed at WIYN have had separations below the diffraction limit of $\sim$50 mas
at 692 nm
(Horch \ea 2011b), which has led to
follow-up observations at larger telescopes and the determination of
a handful of short-period orbits, with periods of a few months to a few years (Horch \ea 2015b). 
In the area of exoplanet host star observations, we have 
developed a methodology for companion detection from reconstructed images
obtained from speckle data (Horch \ea 2011a, Howell \ea 2011), and used this
to estimate the fraction of {\it Kepler} exoplanet hosts that also have a 
stellar companion (Horch \ea 2014). We have also set limits on
stellar companions for exoplanets with eccentric orbits (Kane \ea 2014)
and developed techniques for assessing whether a companion star is likely
to be gravitationally bound to its primary (Everett \ea 2015; Teske \ea 2015).

The observations described here 
represent our seventh (and so far 
largest) installment of relative astrometry and photometry of binary stars. 
This brings the total number of measures contributed by the 
WIYN speckle program
to over 8000. We anticipate that after this there will be one more large group 
of measures to be published from data taken at WIYN on the general binary
survey work
that will cover the time frame of mid-2012 to the end of 
2013. From that point forward, the observing lists changed significantly to
accommodate new scientific goals including more exoplanet host star 
observations, and less time was spent on binary star surveys. Most {\it
Hipparcos} double stars and suspected doubles within 250 pc of the solar
system and observable from WIYN 
had been measured by our program by the end of 2013, and 
many of the binaries amenable to providing astrophysical information
within the next decade had been identified. Since that point, the focus
has been on understanding multiplicity as a function of spectral type
and metallicity, and in following up on promising systems for mass estimates. 
This paper, while mainly
contributing to the earlier survey work, also points the direction
to these new projects involving star formation and multiplicity statistics by
providing first orbits for three K-dwarf systems.

\section{Observations}

The observations were carried out over six runs at the WIYN telescope,
specifically, 
17-21 Sep 2010, 
23-26 Oct 2010,
11-16 Jun 2011,
7-11 Sep 2011, 
10-11 Dec 2011, and
4-8 Feb 2012.
In each case, an observing list was constructed primarily from 
{\it Hipparcos} Double Stars (HDSs) and {\it Hipparcos} suspected 
doubles (ESA 1997), double-lined spectroscopic binary stars identified 
in the Geneva-Copenhagen spectroscopic survey (Nordstr\"{o}m \ea
2004), and stars we have previously found to be double in our 
own program and reported in earlier papers in this series.

For all observations here, the Differential Speckle Survey Instrument (DSSI)
was used (Horch \ea 2009). The instrument can mount to either of the Nasmyth 
ports of the WIYN telescope and takes speckle observations in two 
filters simultaneously. 
The DSSI
observing program at WIYN began in 2008, and the instrument was upgraded to use
two electron-multiplying CCD cameras in January of 2010. More recently,
DSSI also been used
at Lowell Observatory's Discovery Channel Telescope (DCT), and at both the
Gemini North and Gemini South telescopes.

The typical observing sequence was to record 1000 40-ms frames
of a target star in two filters simultaneously. Each frame consists of a
$128 \times 128$-pixel subarray on the chip, which represents a field of
view of approximately $2.8 \times 2.8$ arc seconds. This was followed 
by a similar sequence of a bright (5th magnitude)
unresolved star that is within a few degrees of the science target. The
latter serves as an estimate of the speckle transfer function, which 
is then used in the data analysis. For fainter targets, we sometimes 
took more than 1000 frames; for example, for a 12th or 13th magnitude star, we
would typically take 3000-5000 frames, depending on seeing conditions. 
To make observations more efficient, we would observe up to several science
targets that were close together on the sky in a sequence of increasing right
ascension, working in a single bright unresolved calibrator at the appropriate
moment based on its right ascension.  This star was then 
used as the calibrator for all science targets in the group. 
The resulting data files were stored in FITS
format, each object having two files, one for each wavelength used.

\section{Data Reduction and Analysis}

\subsection{Determination of the Pixel Scale and Orientation}

The pixel scale and orientation were determined in a similar fashion as
in our previous work at WIYN. A slit mask that 
mounts to the tertiary baffle support structure on the telescope is used.
Once in place, this allows us to take speckle data of
a very bright unresolved star (typically 2nd or 3rd magnitude)
through the slits. The resulting images show a series of superimposed
fringes of different spatial frequencies. If a speckle sequence is 
taken and the spatial frequency power spectrum of the ensemble of 
images calculated, these fringes map to sharp, linear features 
in the Fourier plane that
can be fitted with high precision. Using a measure of 
the separation of the slits, the
distance from the mask to 
the focal plane, the wavelength of the light used, and the focal ratio
of the telescope, the spatial frequencies obtained from the
fit in terms of cycles per pixel can be translated into cycles per 
arc second. Upon inverting, this then 
yields the scale in arc seconds per pixel.

Three of the four quantities needed for the calibration
are straightforward to measure: observatory
staff have measured the distance from the baffle support to the nominal
focal plane to much better than 0.1\%. The slit spacing has been measured
directly from the mask to a similar precision by one of us (E.H.). 
The focal ratio of the 
telescope is of course well-known. The final quantity, the effective 
wavelength of the observation, is a combination of the filter and 
atmospheric transmission, quantum efficiency of the detector,
and the spectrum of the star being observed. In order to
get the best possible value of this quantity, we use the Pickles (1998) spectral
library, interpolating to the correct spectral type when needed, and we
combine this with vendor-provided curves for quantum efficiency and
filter transmission. A Kitt Peak atmospheric transmission curve is available
from the observatory website. By simulating the wavelength distribution 
that is in fact detected, we can then estimate the effective wavelength
as the average value of the photon flux of this distribution.

As we have reported in earlier papers, the channel of DSSI where light
reflects off of the dichroic element inside the camera (what
we refer to as ``Channel B'') exhibits a slight variation in pixel scale as a
function of position angle. This distortion is consistent with what would 
be expected if a reflective surface in the instrument, such as the 
dichroic, were not oriented at 
exactly at $45^{\circ}$ relative to the incoming optical axis. We have 
extensively mapped out the effect by taking slit mask data at different
orientations of the mask relative to the detector axes. This is easily
accomplished by rotating the instrument adapter subsystem on the Nasmyth
port. Figure 1 shows an example of the results in scale obtained as a
function of rotator offset angle. In the case of the Channel B data, 
the variation is about $\pm 2$\% from the mean, a typical result at WIYN. In contrast, the
Channel A data show no clear evidence of such an effect, and so 
we simply average all of the values obtained for a given run
to obtain the final value of the scale for that channel. The standard
error of these values has an average of 0.026 mas/pixel;
we view this as a lower limit to the uncertainty in the scale, as it is
obtained from a single sequence of observations on only one star.
 
One possible source of systematic error in the scale determination is
the effective wavelength of the slit mask observations, if the filter
transmission curves or the stellar spectrum selected from the Pickles 
library are not good models of the observations. In order to 
investigate the effect of color variation on the scale determination,
we observed two stars with the slit mask in a single run on three occasions.
The stars were chosen to have very different spectral types, either A and M or
A and K. If we subtract the two scales obtained in a given 
channel, the difference should be zero
if there is no systematic effect in color. The value obtained in these
cases is $-0.036 \pm 0.023$ mas/pixel,
showing that 
the systematic error due to the color calibration is therefore at most very modest. 
A conservative total scale uncertainty can be obtained by adding this value in
quadrature with the 0.026 mas/pixel figure from the internal repeatability mentioned
above, resulting in 0.044 mas/pixel, or about 0.2\% of the nominal scale value.
%

The orientation angle between celestial coordinates and pixel axes
is determined mainly by taking 1-s images of stars from {\it The 
Bright Star Catalogue} (Hoffleit and Jaschek 1982), and offsetting the 
telescope in different directions
between exposures. By using the full area of the chip, we can obtain
a sufficient baseline to measure the offset angle
to $\sim$0.5$^{\circ}$ in good conditions for a given run. Such
a sequence of observations is usually taken nightly during the run,
giving several measures per run, but if the sequence is judged to have been
taken in poor seeing or under windy conditions, we sometimes removed a
nightly sequence from further consideration. Using only runs where we have
three or more sequences meeting the ``high-quality'' criteria, 
we conclude that the standard error in offset angle
has average value of 0.38$^{\circ}$ per run when combining information in both 
channels. 

Regardless of the offset angle between pixel axes and celestial coordinates,
Channels A and B should be mirror images of one another about some 
symmetry axis that
is nearly aligned with one of the pixel axes. To establish this small 
internal offset angle, 
we sought to combine both the A and B
results to get a single value of this offset with the lowest random uncertainty.
The mask files can be used in this regard to establish the axis of symmetry
relative to the pixel axes  to high precision
by assuming that any offset angle in Channel B is the same
relative to this axis as in Channel A, although opposite in direction. Thus, if the 
orientation angles obtained are merely averaged, then
the result is an estimate of the angle made between the axis of symmetry 
and the pixel axes. For each run, we determined the mean value of this angle, and
then averaging these results over all runs, we find a value of 
$\theta_{0} = 0.12 \pm 0.13^{\circ}$. This figure is used with all runs in our analysis here.
Then, the offset angle between this symmetry axis and celestial coordinates
in Channel A is given by $+\theta_{1}$, and $-\theta_{1}$ in Channel B, where
$\theta_{1}$ is the average of the absolute value of the offset obtained 
using the 1-s offset images. 
In this way, the angle
between the symmetry axis and the pixel axes remains the same from run to
run, as it is most likely internal to the DSSI instrument, 
but the size of the offset in both directions relative to that axis
varies from run to run, as that is most likely related to the small
variations in the zero point of the 
instrument rotator and the mounting of the instrument on the telescope,
which may change from
run to run.
Given this, the total uncertainty for the angle 
is obtained by adding the uncertainties in the determination of the symmetry
axis and that of the 1-s images; we find $\sqrt{0.38^{2} + 0.13^{2}} = 
0.40^{\circ}$ as the final angle uncertainty from this calibration. This is
assumed for all runs, as shown in Table 1.


\subsection{Speckle Data Reduction}

While our reduction method from raw speckle frames remains the same as in
previous papers in this series, 
a brief summary is provided here for convenience. 
The frames are de-biased and the autocorrelation of each frame 
in the observation is computed.
These are then summed. The same computation is performed on the speckle
frames of the calibration point source. Both autocorrelations are 
Fourier transformed, and then the result for the science target is divided by that of 
the point source to deconvolve the speckle transfer function
from the observation. The square-root of this function is 
then computed, which forms the diffraction-limited modulus estimate for 
the science target's Fourier transform.

Lohmann \ea (1983) first described how phase information of the object's 
Fourier transform is contained in the summed triple correlation function of speckle images, and
its Fourier-transform pair, the image bispectrum. We compute near-axis sub-planes of the 
bispectrum are then computed for the science observation following their 
methodology. The phase
of the object's Fourier transform is then computed from these, 
using the relaxation algorithm of Meng \ea (1990). This is combined
with the modulus estimate previously obtained to give the (complex)
Fourier transform of the object. This is low-pass filtered with a 
2-dimensional Gaussian function, and inverse transformed. The 
result is a diffraction-limited reconstructed image of the object.
This is used to identify secondary and tertiary components
 in the case of binaries and
trinaries, and used in the case of non-detections to estimate the 
detection limit of the observation as described further in Section 3.4.

In the case of binary and trinary stars, once rough positions of the 
components are determined from the reconstructed images, these 
are used as input for a fitting algorithm that computes final 
position angles, separations and magnitude differences of the components.
This is a Fourier-based routine; the power spectrum of the object is
formed by Fourier transforming the autocorrelation, and then it is divided
by the point source power spectrum. The result for a binary star
would be a pure fringe pattern, related to the square of a 
two-dimensional cosine function. From the separation, orientation,
and depth of the fringes, the relative astrometric and photometric 
parameters for the components
can be determined. We use a weighted least-squares fit to the power spectrum
in order to arrive at the final parameters for each observation. 
The methodology of the weighting dates back to Horch \ea (1996), where
as described there, we attempt to approximate a $\chi^{2}$-minimization
procedure. In order to do that, one must understand the true noise 
statistics in the power spectrum, which in turn requires a knowledge of
the read noise of the detector and the number of photons per ADU as a 
function of the electron multiplying gain.
We have studied this with our EMCCDs and built this into the 
current weighting 
model. More recently, Pluzhnik (2005) has provided a rigorous method for applying similar
ideas to speckle data even for non-circular speckle transfer functions; 
our method is more heuristic in the sense that the variance of points
in the power spectrum
is estimated from the (properly scaled) data themselves and the final variance
needed for calculating $\chi^2$ is not an analytical function but instead 
a smoothed version of the derived variance function
in the Fourier plane.

\subsection{Double Star Measures}

Our measures of double stars are found in Table 2. The columns give:
(1) the Washington Double Star (WDS) number (Mason \ea 2001), which also
gives the right ascension and declination for the object in 2000.0 coordinates;
(2) the Aitken Double Star (ADS) Catalogue number,
or if none, the Bright Star Catalogue ({\it i.e.\ }Harvard Revised [HR]) number,
or if none, the Henry Draper Catalogue (HD) number,
or if none the Durchmusterung (DM) number of the object;
(3) the Discoverer Designation;
(4) the {\it Hipparcos} Catalogue number;
(5) the Besselian date of the observation;
(6) the position angle ($\theta$) of the
secondary star relative to the primary, with North through East defining the
positive sense of $\theta$; (7) the separation of the two stars ($\rho$), in
arc seconds; (8) the magnitude difference ($\Delta m$)
of the pair in the filter used;
(9) the center wavelength of the filter; and
(10) the full width at half maximum of the filter transmission.
The position angle measures have not been precessed from the dates shown.
Sixty-six pairs in the table have no 
previous detection of the companion
in the {\it Fourth Catalogue of 
Interferometric Measures of Binary Stars} (Hartkopf,
\ea 2001a); we propose discoverer designations of
YSC (Yale-Southern Connecticut) 168-231 here. (Two systems discovered
were trinaries.) This continues the collection
of YSC discoveries detailed in our earlier papers in this series.
In Figure 2, we show two characterizations of the data set as a whole; 
in panel (a) we plot the magnitude difference obtained as a function of
separation, while in panel (b) we plot the same as a function of system
apparent $V$ magnitude.

\subsubsection{Astrometric Accuracy and Precision}

The astrometric accuracy of the data set is important to establish 
so that, when used in orbit calculations, a proper weighting can
be used. Since the DSSI camera is a two-channel instrument,
a fundamental calibration in this regard is a comparison between the
results obtained in the channels for each observation. This is 
shown in Figure 3 for both position angle and separation. All observations
in Table 2 are included. The results show that the
repeatability between the two channels in position angle is a function
of separation, while the repeatability in separation is essentially independent
of separation. The average difference between the channels in position angle
is $-$0.01 $\pm 0.04$ degrees, with a standard deviation of $1.51 \pm 0.03$
degrees. In separation, the average difference is $0.06 \pm 0.07$ mas, with
a standard deviation of $2.45 \pm 0.05$ mas. As these standard deviation
values are the result of the subtraction of two measures that may
be assumed to have similar precisions, the precision of individual measures
in a single channel can be estimated by dividing these results by
a factor of $\sqrt{2}$, resulting in an overall internal precision of
$1.07 \pm 0.02$ degrees and $1.73 \pm 0.04$ mas. These numbers set 
the lower limit of the uncertainty for the data set. If 
the astrometry of paired measures from the two channels are averaged (which
we did not do in Table 2), then there would be a further reduction of
a factor of $\sqrt{2}$ in that case, resulting in a separation repeatability
of about 1.2 mas for example, which is comparable to previous data sets 
we have analyzed from WIYN.

The position angle repeatability is expected to be a function of separation.
Position angle measures may be viewed as having a linear measurement
uncertainty equivalent to that of the separation measures, but orthogonal
in direction ({\it i.e.\ }orthogonal to the vector direction of the separation). In that
case, the position angle uncertainty is equal to $\delta \rho/\rho$ in 
radians, or using the value for $\delta \rho$ obtained here of 1.73 mas
and converting to degrees, this is $0.099 / \rho$, where $\rho$ is 
entered in arc seconds. This is roughly consistent with Figure 3(a).

To determine the accuracy of our measures, we examine the results obtained
here for stars that already have well-determined orbits in the literature;
specifically, we use systems that have either Grade 1 or Grade 2 orbits
in the {\it Sixth Catalog of Visual Binary Star Orbits} (Hartkopf \ea 2001b). 
Only systems that have uncertainty estimates to the
orbital elements are considered, so that we can 
propagate errors to estimate the uncertainties in the ephemeris positions
and that the ephemeris uncertainties are less than 4$^{\circ}$ and 8 mas
respectively. These limits were chosen so that the orbits are of high
quality, but also so that the ensemble of systems of sufficient size to 
be statistically meaningful.
We then calculate the ephemeris position angle and separation for the system
for the epoch of our observations, and use the measures in Table 2 to 
obtain an observed minus ephemeris residual in both $\rho$ and $\theta$. The 
systems used for this study appear in Table 3. The residual plots in 
position angle and separation are shown in Figure 4. The repeatability 
uncertainties for single observations
derived above, namely $0.099/\rho$ in position angle and 1.73 mas in
separation, are plotted as solid lines. Also plotted as dotted lines are the uncertainties
stated in the previous section for the measurement of the pixel scale
and orientation. The plots indicate that, for position angle, the 
internal repeatability dominates the overall uncertainty for separations
under 0.3 arc seconds, and above that
value, the measures are limited by the orientation measures. 
For the separation measures, uncertainties for separations under $\sim$0.85 arc 
second are dominated by
the internal repeatability while larger separations are limited by the
the scale measurement. In searching for systematic error by studying
average residuals, 
we find an average offset in position angle in Figure 4(a) of
$-0.5 \pm 0.4^{\circ}$ for the entire sample and $-0.3 \pm 0.4^{\circ}$ 
for the sub-sample of systems with ephemeris uncertainties of less
than 2$^{\circ}$. For separation, the entire sample has an average 
residual of $-0.9 \pm 0.4$ mas, with the subset of measures with 
ephemeris uncertainties of less than 4 mas yielding $-0.6 \pm 0.4$ mas.

Two stars with high-quality orbits in the {\it Sixth Orbit Catalog} 
were not included for the study here. 
The first is FIN 312 = HIP 12390 = HR 781 = 
$\epsilon$ Cet. Although there is a Grade 1 orbit in the literature 
calculated by Docobo \& Andrade (2013), measures published 
in 2014 and 2015 have all shown a 
slightly smaller separation than predicted by 
about 3 mas, and most since 2005 have the same offset,
although the effect appears smaller in the 2005-2013 timeframe.
The second 
is STF 1670 = HIP 61941 = $\gamma$ Vir. 
This well-known visual binary has a Grade 2 orbit calculated
by Scardia \ea in 2007, but has 
trended toward smaller separations than predicted since that time;
the system has recently passed through periastron and is now rapidly
increasing in separation.
It is also the case that, at a separation 
of 1.65 arcseconds, the uncertainty in our plate scale 
determination dominates over the random uncertainties 
of internal measurement precision. So, 
our measures at that separation should be considered 
less reliable.

Overall, we conclude that there is no evidence of significant 
systematic offsets
in either position angle or separation. At worst, there is a modest
offset in separation of a fraction of 1 mas, but this could easily be
accounted for in the uncertainty of the orbital elements used and the 
relatively small number of systems that meet our quality criteria. 
Therefore
a good measure of the total uncertainty for any measure in Table 2 is 
obtained by adding the orientation and scale uncertainties in quadrature
with the internal random uncertainties obtained from the repeatability
study. For example, a system with a separation of 0.1 arc seconds = 100 mas would have 
a position angle uncertainty of $\sqrt{0.40^2 + (0.099/0.1)^2} = 1.1^{\circ}$ and 
a separation uncertainty of $\sqrt{(0.044/22)\times100)^2 + 1.73^2} = 1.74$ mas,
whereas for a system with a separation of 1.0 arc second = 1000 mas, the results would be
 $\sqrt{0.40^2 + (0.099/1.0)^2} = 0.41^{\circ}$ degrees in position angle and
 $\sqrt{(0.044/22)\times1000)^2 + 1.73^2} = 2.64$ mas in separation.

\subsubsection{Photometric Precision}

To judge the photometric precision, we first compute the parameter that
in previous papers we have called $q^{\prime}$, which is given by the
seeing of the observations multiplied by the separation of the pair. 
As shown for example in Horch \ea (2009), this parameter, in arc seconds
squared, should be proportional to the separation divided by the 
isoplanatic angle, therefore providing a measure of to what degree
(in relative terms) the observation is isoplanatic. If an observation
lacks isoplanicity, one would expect that the magnitude difference obtained
from a speckle analysis would be systematically too large. 

In Figure
5, we show the differences in $\Delta m$ obtained from our measures in 
Table 2 versus the {\it Hipparcos} measures, $\Delta H_{p}$. Giants, variable stars, and
trinary systems are removed from the sample prior to plotting, as are
stars with $B-V$ colors greater than 0.6.
None of our filters match
the {\it Hipparcos} $H_{p}$ filter, but it is instructive to see the 
comparison with each of the filters we used. The closest match is our
562-nm filter, which has width $\Delta \lambda = 40$ nm. We see in 
Figure 5(a) that the average difference $\Delta m - \Delta H_{p}$ is modest, 
and that the largest differences are found at the highest values of
seeing times separation, above approximately 0.55 arcsec$^{2}$. In
the case of the redder filters (panels c and e of the figure), a
negative difference is present at low values of $q^{\prime}$, but
above 0.55 arcsec$^{2}$, again we see a rough trend toward large
positive offsets. This is expected, since in cases of high isoplanicity
(and low $q^{\prime}$), redder filters would generate a smaller magnitude difference than 
the bluer $H_{p}$ filter for main-sequence stars, given the intrinsic 
color differences of the companions. However, for high values of 
$q^{\prime}$, the lack of isoplanicity eventually overwhelms that
effect and leads to a larger $\Delta m$ regardless of the color difference.
In these panels,
the data points are divided into two sub-groups depending on the 
estimated uncertainty of the $\Delta H_{p}$ measure in the {\it Hipparcos}
Catalogue. If less than 0.1 magnitudes, then the point is plotted as 
a filled black circle, 
and if greater, then as an open circle with the color
indicating the filter.

In panels (b), (d), and (f) of the figure, we plot the $\Delta H_{p}$
value as a function of our $\Delta m$ for the high-quality subset of
measures from panels (a), (c), and (e) respectively. Specifically,
these are the measures that have $\delta(\Delta H_{p}) < 0.1$ magnitudes
and $q^{\prime} < 0.55$ arcsec$^{2}$.
For the 562-nm filter, which is the closest match to $\Delta H_{p}$, 
we find  an average offset for this subset of $0.01 \pm 0.03$ magnitudes,
with a standard deviation of $0.12 \pm 0.02$ magnitudes. Some of this
uncertainty is due to the {\it Hipparcos} measure; the average uncertainty
in this sub-sample is 0.052 magnitudes. Subtracting this value in quadrature
from the standard deviation of the differences, we obtain 0.10 magnitudes.
Thus, this is the best measure of the uncertainty of our magnitude differences.
The other filters show larger deviations from the unitary line and with larger 
scatter, as expected in the case of a sample of main sequence stars with
a variety of colors. This should not be interpreted as a decrease in photometric
precision in these cases, but merely the result of the difference in central 
wavelength of the filters. We also examined cases in Table 2 where four or more 
magnitude difference measures appear for a given pair in a single filter, and we computed the
standard deviation of each group of observations. The average of these 12 cases,
which are mainly observations in the 692-nm filter, is
$\langle\sigma(\Delta m)\rangle = 0.09 \pm 0.02$ magnitudes, in good agreement with
the uncertainty derived from the above study.
Therefore, we believe the result of uncertainties typically
$\pm0.1$ obtained in the 562-nm filter is probably indicative of the other filters as well.

\subsection{Non-detections}

In a number of cases, we observed stars under good conditions and failed to
detect a companion star. The majority of these cases are examples of either
{\it Hipparcos} suspected double stars or stars found to be double-lined
spectroscopic binary stars by the Geneva-Copenhagen spectroscopic survey
of G-dwarfs. (For many stars in the latter category, the separation is most likely
too small to be measurable 
from WIYN.) For these cases, we have derived a detection limit estimate
as a function of separation using the method described in previous papers.
Briefly, we examine annuli in the reconstructed images that are centered
on the central star and have center radius of a desired value. We 
determine all local maxima and minima in the annulus, and derive their 
mean value and standard deviation. We then estimate the detection limit
as the mean value of the maxima plus five times the average sigma of the 
maxima and minima. 

The current reduction pipeline produces a curve of this detection limit
as a function of separation. Examples are shown in Figure 6. The typical
curve rises sharply from the diffraction limit to a ``knee'' at 
a separation of 0.15-0.2
arc seconds, then it flattens out somewhat but continues to increase out to
the outer limit of the plot, which is 1.2 arc seconds. 
It is interesting to note that this curve is very similar to the envelope of points
in Figure 1(a), which is essentially a plot of all detected companions
with the same axes.
The degree of the 
flattening above 0.2 arc seconds
is mainly determined by signal-to-noise ratio and how well
the point source chosen for our Fourier deconvolution matches that of the
science target, with high signal-to-noise and good matches leading to less
flattening, while poor signal-to-noise and/or less perfect matches leading
to more flattening. Thus, the curve can be roughly characterized by its
value at two points: the knee, and the largest separation point. The
curve can then be roughly reconstructed by drawing a line from a magnitude
difference of zero at the diffraction limit up to the knee, and a second line
from the detection limit at the knee to that at a large separation. 
In Table 4, we show detection limits determined in this way for 0.2 
and 1.0 arc seconds for 176 stars for which no companions were found.

\section{Orbits for Three Nearby K-Dwarf Binaries}

Much of our binary star speckle work at WIYN to date has been focussed on {\it Hipparcos}
double stars within 250 pc of the Solar System, without regard to 
spectral type. However, there are compelling reasons to refocus attention
on a sample of K-dwarfs that are nearby. 
For G-dwarfs and M-dwarfs, multiplicity studies have either been completed or started
(Raghavan \ea 2010; Winters \ea 2015). However, for K-dwarfs, less work 
has been done and yet
for the nearest systems, speckle
imaging samples separations that are comparable to the scale of our own
Solar System.
The main studies enabled by a target list of K-dwarfs are therefore to
(a) discover how unusual our Sun is in its stellar solitude, 
(b) understand how many
stars of different types are multiples, 
(c) provide fundamental statistics that will drive
theoretical work in the area of star formation, and 
(d) provide a list of stars where nearby analogs to the Solar
System might be found because they lack stars on planetary formation scales.

Because of these facts, two of us (T.H. and J.W.) identified a
sample of K-dwarfs within 50 pc of the solar system using the
RECONS\footnotemark \hspace{4pt}25-pc database and {\it Hipparcos}
results to 50 pc, many of which 
had not been previously observed with speckle imaging. Although we have
begun to systematically observe these stars primarily at the DCT and
Gemini, it was also useful to look at which of the stars presented
here were serendipitously stars on this list.
A total of 48 observations in Table 2 were made of 15 targets on the list of K-dwarfs, 
including three systems resolved here for the first time:
YSC 198, 206, and 208. 
YSC 198 is actually the inner component of a triple star system, with the tertiary
component having separation of $\sim$5 arc seconds.
Four stars on the K-dwarf list were unresolved
and appear in Table 4.
Of all the K-dwarfs observed, three systems had sufficient data in 
the literature so that, combined with the astrometric data presented here,
we were able to attempt first orbit calculations. We discuss
these systems below, and orbital parameters are found in Table 5.
Interestingly, two of these three systems have a previously-known wider component.

\footnotetext{http://www.recons.org/}

\subsection{HDS 99Aa,Ab}

This K7V pair at 31 pc has an orbital period of 8.7 years and
semi-major axis of 132.1 mas, which translates to a physical separation of
4.2 AU. 
The orbit we calculated did not use the {\it Hipparcos} data point, due to the 
relatively short period compared with the length of the {\it Hipparcos}
mission. The remaining data nonetheless span nearly a full
period since 2007. The magnitude difference of the pair is less than 1, 
so we may estimate that the pair consists of perhaps a K7 primary
star with a K9 secondary; this would imply a mass sum in the range of 
1.15$M_{\odot}$, using a standard reference such as Schmidt-Kaler (1982). 
This is consistent with the mass sum calculated from the
orbital parameters shown in Table 5, which already has a relatively small
uncertainty.
This system forms the inner pair of the common proper motion double LDS 3195AB,
which has separation of 141 arc seconds (or 4450 AU).
Our orbit is shown in Figure 7.

\subsection{WSI 74Aa,Ab}

The SIMBAD\footnotemark \hspace{4pt}database shows the spectral type of this pair to 
be K2.5V, which forms the
inner system of HDS 1795AB. Given the magnitude difference of approximately
1.4, we estimate that this is a K2V+K6V system, implying a mass sum of 
1.38$M_{\odot}$, based again on data in Schmidt-Kaler (1982). 
While the orbital elements do not permit anything more than a very rough
mass sum to be inferred at this point, it is nonetheless consistent with this value. We
find a period of 2.6 years and a semi-major axis of 88 mas (2.1 AU). The
wider component, which was out of our small field of view, is at a projected
separation of 2.7 arc seconds (64 AU).
The 4th Interferometric Catalog contains several non-detections by Tokovinin, but these are at
or below the diffraction limit during the epochs in question
for the orbit presented here. The plot of the orbit is shown in Figure 8.

\footnotetext{http://simbad.u-strasbg.fr/simbad}

\subsection{HDS 2053}

This system has the largest magnitude difference of the three systems
under consideration here at approximately 2.3 magnitudes, so despite
relatively few data points to work with for the orbit calculation, the determination
of the quadrant for each observation should not be in question. We calculated
this orbit with and without the {\it Hipparcos} point, given the orbital 
period. We present elements without the {\it Hipparcos} point in Table 5; when
the point is included, the period is shorter by approximately 4 years, but
the semi-major axis is similar. However, looking at the position of the
{\it Hipparcos} point on the orbit, it is clear that there is a potential
to derive a systematically low separation if averaging the position of the
secondary over the lifetime of the {\it Hipparcos} mission. The resulting total mass
estimate is $1.5 \pm  0.5 M_{\odot}$. We estimate that this system consists of
a K4V primary star with a M1V secondary, so that the expected total mass
would be $\sim 1.2 M_{\odot}$. The semi-major axis that we derive 
using the revised {\it Hipparcos} parallax is 9.5 AU. We show our orbit
in Figure 9.

\section{Summary}

We have presented the results of over 2800 observations of double stars
and suspected double stars using the dual-channel speckle imaging camera,
DSSI, when it was resident at the WIYN Observatory at Kitt Peak during
2010-2012. The astrometric precision appears to be 
in line with previous papers in this series, generally $\sim$2 mas in 
separation and $\sim (0.1/\rho)^{\circ}$ in position angle, where 
$\rho$ is entered in arc seconds. The photometric precision
is generally about 0.1 magnitudes per observation. In cases where
there is evidence for a lack of isoplanicity or if the secondary falls
near the edge of the frame, we have not reported magnitude differences.
One hundred seventy six additional objects showed no evidence of a
companion and in these cases we have derived 5-$\sigma$ detection limits
at separations of 0.2 and 1.0 arc seconds.

The data presented here, along with existing relative astrometry already
in the literature, permitted the calculation of preliminary orbits
for three K-dwarf binaries at distances of 24 to 42 pc. 
The periods ranged from 2.6 to 23.5 years,
and initial mass estimates from the orbital elements are consistent with
the expected theoretical values for stars of that spectral type. 
These serve as examples for the potential
of sustained, well-calibrated astrometry efforts on such systems.

\acknowledgments

The authors would like to thank all of the excellent staff at the WIYN
telescope for their help during our observing runs. We were privileged to
work with professionals of such dedication and skill.
We used the SIMBAD database, 
the Washington Double Star Catalog, the Fourth
Catalog of Interferometric Measures of Binary Stars, and the Sixth Orbit
Catalog in the preparation of this paper.
We gratefully acknowledge the role of the Kepler Science Office 
in upgrading DSSI
to the two-EMCCD mode used here, support from 
National Science Foundation
grant AST-0908125, which funded the observations discussed here, and grant 
AST-1517824,
which funded the completion of the analysis and publication of this work.


\clearpage


\begin{centering}
\begin{deluxetable}{lcccccc}
\tabletypesize{\scriptsize}
\tablewidth{0pt}
\tablenum{1}
\tablecaption{Pixel Scale, Orientation, and Filters Used on Each Run}
\tablehead{ 
&
\multicolumn{3}{c}{Channel A} &
\multicolumn{3}{c}{Channel B}  \\
\colhead{Run} & \colhead{Offset Angle} &
\colhead{Scale} &
\colhead{$\bar{\lambda}$} &
\colhead{Offset Angle} &
\colhead{Scale} &
\colhead{$\bar{\lambda}$} \\
 & \colhead{($^{\circ}$)}
 & \colhead{(mas/pix)} & \colhead{(nm)}
 & \colhead{($^{\circ}$)}
 & \colhead{(mas/pix)} & \colhead{(nm)}
} 
\startdata 
Sep 2010 & $+5.51$ & $21.793$ & 692 & $-5.27$ & 22.447 & 562,880\\
Oct 2010 & $+5.89$ & $21.642$ & 692 & $-5.64$ & $22.987$ & 880 \\
Jun 2011 & $+5.99$ & $21.641$ & 692 & $-5.75$ & 22.939 & 880 \\
Sep 2011 & $+5.43$ & $21.807$ & 692 & $-5.19$ & 23.137 & 880 \\
Dec 2011 & $+5.00$ & $21.529$ & 692 & $-4.76$ & $19.812$\tablenotemark{a} & 562 \\
Feb 2012 & $+5.99$ & $21.471$ & 692 & $-5.74$ & 22.792 & 880 \\
\hline
{\it Est. Uncertainty} & $\pm 0.40$ & $\pm 0.044$ & ... & $\pm 0.40$ & $\pm 0.044$ & ... \\
\enddata 
\tablenotetext{a}{A different optical configuration was used in this channel 
compared to the other runs.}
\end{deluxetable}

\clearpage

\pagestyle{empty}
\begin{deluxetable}{lllrcrrrrlc}
\tabletypesize{\scriptsize}
\tablenum{2}
\tablecaption{Double star speckle measures}
\tablehead{
\colhead{WDS} &
\colhead{HR,ADS} &
\colhead{Discoverer} & HIP &
\colhead{Date} &
 \colhead{$\theta$} & \colhead{$\rho$} &
\colhead{$\Delta m$} &
\colhead{$\lambda$} &
\colhead{$\Delta \lambda$} & \\
\colhead{($\alpha$,$\delta$ J2000.0)} &
\colhead{HD, or DM} & \colhead{Designation} &&
\colhead{($2000+$)} &
\colhead{($^{\circ}$)} &
\colhead{(${\prime \prime}$)} &
 & \colhead{(nm)} &
\colhead{(nm)} &
}
\startdata

$00022-2052$ & HD 224961  &  YSC 168 &     170 & 
   11.6837 &  39.8 & 0.1105 & 0.42 & 692 & 40 \\
&&&&
   11.6837 &  40.2 & 0.1156 & 0.88 & 880 & 50 \\
&&&&
   11.9430 &  39.9 & 0.1101 & 0.44 & 692 & 40 \\
&&&&
   11.9430 &  37.0 & 0.1176 & 0.65 & 562 & 40 \\

$00031+5228$ & HD 225064 & HDS 3 &     250 & 
   11.9431 & 333.5 & 0.3221 & 3.16 & 692 & 40 \\
&&&&
   11.9431 & 333.5 & 0.3184 & 2.89 & 562 & 40 \\

$00061+0943$ & HD 126 & HDS 7 &     510 & 
   10.7171 & 177.8 & 0.1871 & 0.21 & 692 & 40 \\
&&&&
   10.7171 & 178.0 & 0.1883 & 0.20 & 880 & 50 \\
\enddata

\tablenotetext{a}{Quadrant ambiguous.}
\tablenotetext{b}{Quadrant inconsistent with previous measures in the 4th
Interferometric Catalog.}
\tablenotetext{c}{These measures were previously published in Horch
{\it et al.\ }2011b; this reanalysis should replace the previous values.}
\tablenotetext{d}{These measures were previously published in Appourchaux 
{\it et al.\ }2015; this reanalysis should replace the previous values.}
\tablenotetext{K}{This star is a member of the nearby K-dwarf sample
discussed in the text.}

\end{deluxetable}

\begin{deluxetable}{clrcrrrrl}
\tabletypesize{\tiny}
\tablewidth{0pt}
\tablenum{3}
\tablecaption{Ephemeris Positions and Residuals Used in the Astrometric Accuracy Study}
\tablehead{ 
\colhead{WDS} &
\colhead{Discoverer} &
\colhead{HIP} &
\colhead{Date} &
\colhead{$\theta_{eph}$} &
\colhead{$\rho_{eph}$} &
\colhead{${\Delta \bar{\theta}}$} &
\colhead{${\Delta \bar{\rho}}$} &
\colhead{WDS Orbit Grade} \\
& \colhead{Designation} &&
\colhead{($2000+$)} &
\colhead{($^{\circ}$)} &
\colhead{($^{\prime \prime}$)} &
\colhead{($^{\circ}$)} &
\colhead{(mas)} & \colhead{and Reference}
}
\startdata
$01297+2250$ & A 1910AB &  6966 & 10.7173 & 
  $184.2 \pm 2.4$ & $0.1846 \pm 0.0058$ & $+1.1$ & $-5.5$ & 2, Hartkopf \ea 1996 \\

$02020+7054$ & BU 513AB &  9480 & 10.7145 &
   ($301.6 \pm 4.9$) & $0.6496 \pm  0.0041$ & ($+1.0$) & $-2.7$ & 2, DeRosa \ea 2012 \\
             &          &  9480 & 11.6839 &
   ($306.3 \pm 4.9$) & $0.6299 \pm  0.0042$ & ($+0.9$) & $-2.9$ & 2, DeRosa \ea 2012 \\
             
$02022+3643$ & A 1813AB &   9500 & 10.7200 &
    $356.6 \pm 0.9$ & $0.1647 \pm 0.0017$ & $-0.2$ & $-3.1$ & 2, Hartkopf \ea 2000 \\
             
$02157+2503$ & COU 79  &   10535 & 10.7174 &
    $60.7 \pm 0.3$ & $0.1038 \pm 0.0016$ & $+0.5$ & $-1.8$ & 2, Muterspaugh \ea 2010b \\
             &         &   10535 & 11.6893 &
    $52.2 \pm 0.1$ & $0.1600 \pm 0.0013$ & $+0.4$ & $-0.6$ &  \\

$02262+3428$ & HDS 318  &  11352 & 10.7200 &
   ($160.7 \pm 5.2$) & $0.0928 \pm 0.0060$ & ($-4.3$) & $-4.8$ & 2, Balega \ea 2005 \\ 

$02366+1227$ & MCA  7   &  12153 & 12.0962 & 
  ($326.5 \pm  8.1$)\tablenotemark{a} & $0.0781 \pm 0.0059$ & ($+0.7$) & $-6.1$ & 1, Docobo \ea 2016 \\
\enddata

\tablenotetext{a}{Figures in parentheses did not meet the criteria explained in the text for comparison.}

\end{deluxetable}

\pagestyle{empty}
\begin{deluxetable}{lrcccccccl}
\tabletypesize{\tiny}
\tablewidth{0pt}
\tablenum{4}
\tablecaption{5-$\sigma$ Detection Limits for Unresolved Stars}
\tablehead{
\colhead{WDS} & \colhead{HIP} &
\colhead{Date} &
\multicolumn{3}{c}{5-$\sigma$ Det. Lim., Ch. A} &
\multicolumn{3}{c}{5-$\sigma$ Det. Lim., Ch. B} &
\colhead{Comment} \\
\colhead{($\alpha$,$\delta$ J2000.0)} & &
\colhead{(2000+)} &
\colhead{$\lambda$ (nm)} &
\colhead{$0.2^{\prime \prime}$} &
\colhead{$1.0^{\prime \prime}$} &
\colhead{$\lambda$ (nm)} &
\colhead{$0.2^{\prime \prime}$} &
\colhead{$1.0^{\prime \prime}$} &
}
\startdata

$00043+4007$ &    344 & 10.7143 & 692 & 2.95 & 6.52 & 562 & 3.84 & 5.67 & HIP suspected double. \\
$00302+5428$ &   2364 & 10.7143 & 692 & 3.66 & 4.51 & 562 & 3.25 & 4.38 & HIP suspected double. \\
$00400+1016$ &   3150 & 10.7116 & 692 & 3.02 & 4.80 & 562 & 3.60 & 4.92 & HIP suspected double. \\
$00513-0924$ &   3995 & 10.7172 & 692 & 3.71 & 6.29 & 880 & 3.10 & 5.85 & G-C SB2. \\
             &        & 11.6838 & 692 & 3.55 & 6.74 & 880 & 3.21 & 7.02 &          \\
$00516-0849$ &   4031 & 10.7172 & 692 & 3.89 & 6.54 & 880 & 3.17 & 6.45 & HIP suspected double. \\
             &        & 11.6838 & 692 & 3.64 & 6.98 & 880 & 3.56 & 7.07 &                       \\
$00522-0950$ &   4071 & 10.7172 & 692 & 3.90 & 6.17 & 880 & 3.59 & 5.30 & HIP suspected double. \\
\enddata

\end{deluxetable}

\clearpage

\begin{deluxetable}{lccc}
\tabletypesize{\scriptsize}
\tablewidth{0pt}
\tablenum{5}
\tablecaption{Preliminary Orbits for Three K-Dwarf Systems}
\tablehead{ 
\colhead{Parameter} &
\colhead{HDS 99Aa,Ab} & \colhead{WSI 74Aa,Ab} & \colhead{HDS 2053} 
} 
\startdata 
HIP               & 3493               & 62505             & 71108 \\
Spectrum\tablenotemark{a} & K7V        & K2.5Vk            & K4V  \\
$\Delta m$\tablenotemark{b} & 0.6      & 1.4               & 2.3 \\ 
$\pi$, mas\tablenotemark{c} & $31.81 \pm 2.52$ & $41.96 \pm 3.00$ & $24.28 \pm 2.42$ \\
$P$, years        & $8.667 \pm 0.065$  & $2.643 \pm 0.012$ & $ 23.52 \pm 0.36$ \\
$a$, mas          & $132.4 \pm 1.2$    & $88. \pm 10.$     & $ 230.5 \pm 3.4$ \\
$i$, degrees      & $171.5 \pm 6.6$    & $59.0 \pm 5.7$    & $ 103.10 \pm 0.68$ \\
$\Omega$, degrees & $66. \pm 25.$      & $147.7 \pm 2.2$   & $  52.15 \pm 0.90$ \\
$T_{0}$, years    & $2011.977 \pm 0.023$ & $2016.35 \pm 0.10$  & $ 2004.49 \pm 0.55$ \\
$e$               & $0.493 \pm 0.010$  & $0.520 \pm 0.047$ & $ 0.361 \pm 0.025$ \\
$\omega$, degrees & $ 8. \pm  26.$     & $139. \pm 10.$    & $ 215. \pm 11. $ \\
$M_{\rm{tot}}$, $M_{\odot}$ & $0.96 \pm 0.23$ & $1.31 \pm 0.53$ & $ 1.55 \pm 0.48$ \\
\enddata 
\tablenotetext{a}{As it appears in SIMBAD.}
\tablenotetext{b}{An average of values appearing in the 4th Interferometric 
Catalog that have filter center wavelengths between 500 and 600 nm.}
\tablenotetext{c}{from van Leeuwen (2007).}
\end{deluxetable}

\end{centering}

\clearpage

\begin{figure}[tb]
\plotone{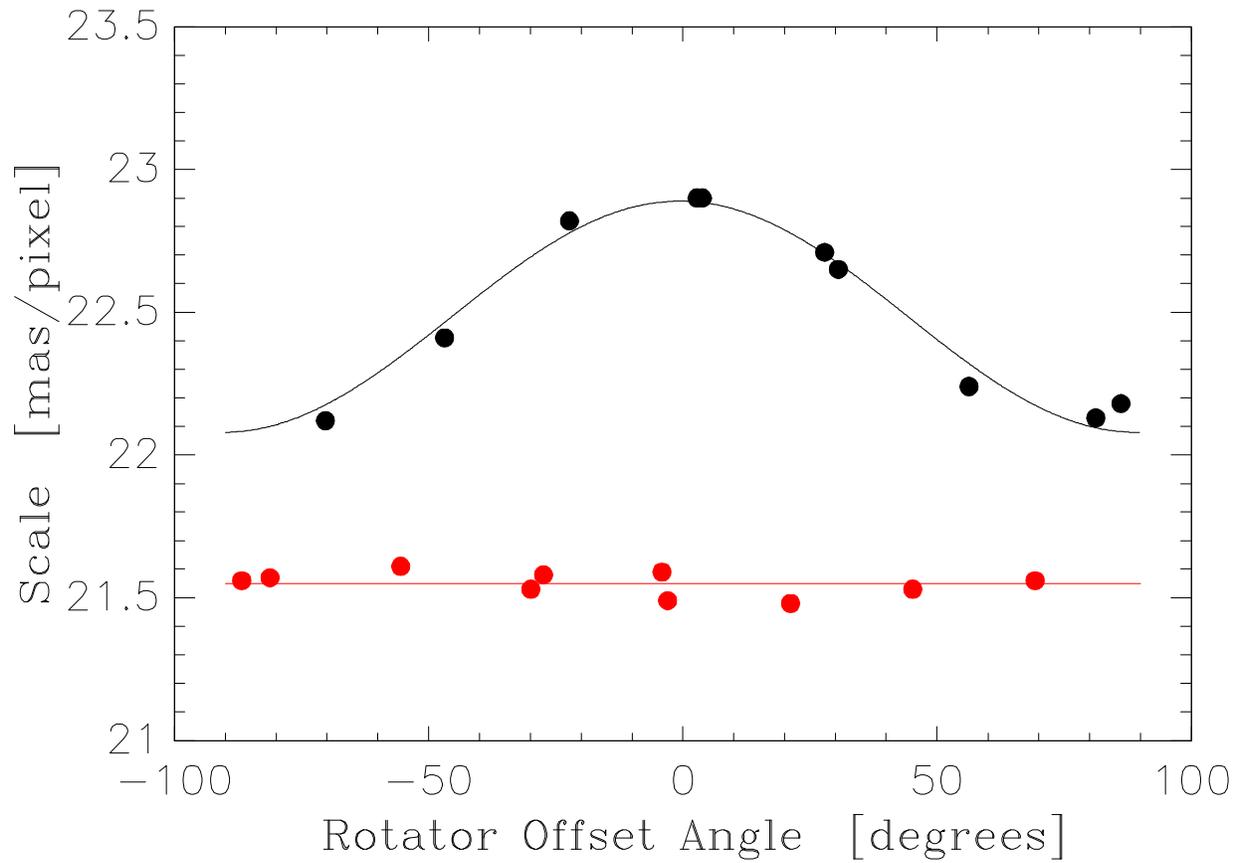}
\caption{
Scale determination as a function of instrument rotator angle. In this
case, HR 5107 was observed through the slit mask in June 2011.
The red points are from Channel A (using the 692-nm filter), and the 
black points are from Channel B (using the 880-nm filter). 
}
\end{figure}

\begin{figure}[tb]
\plottwo{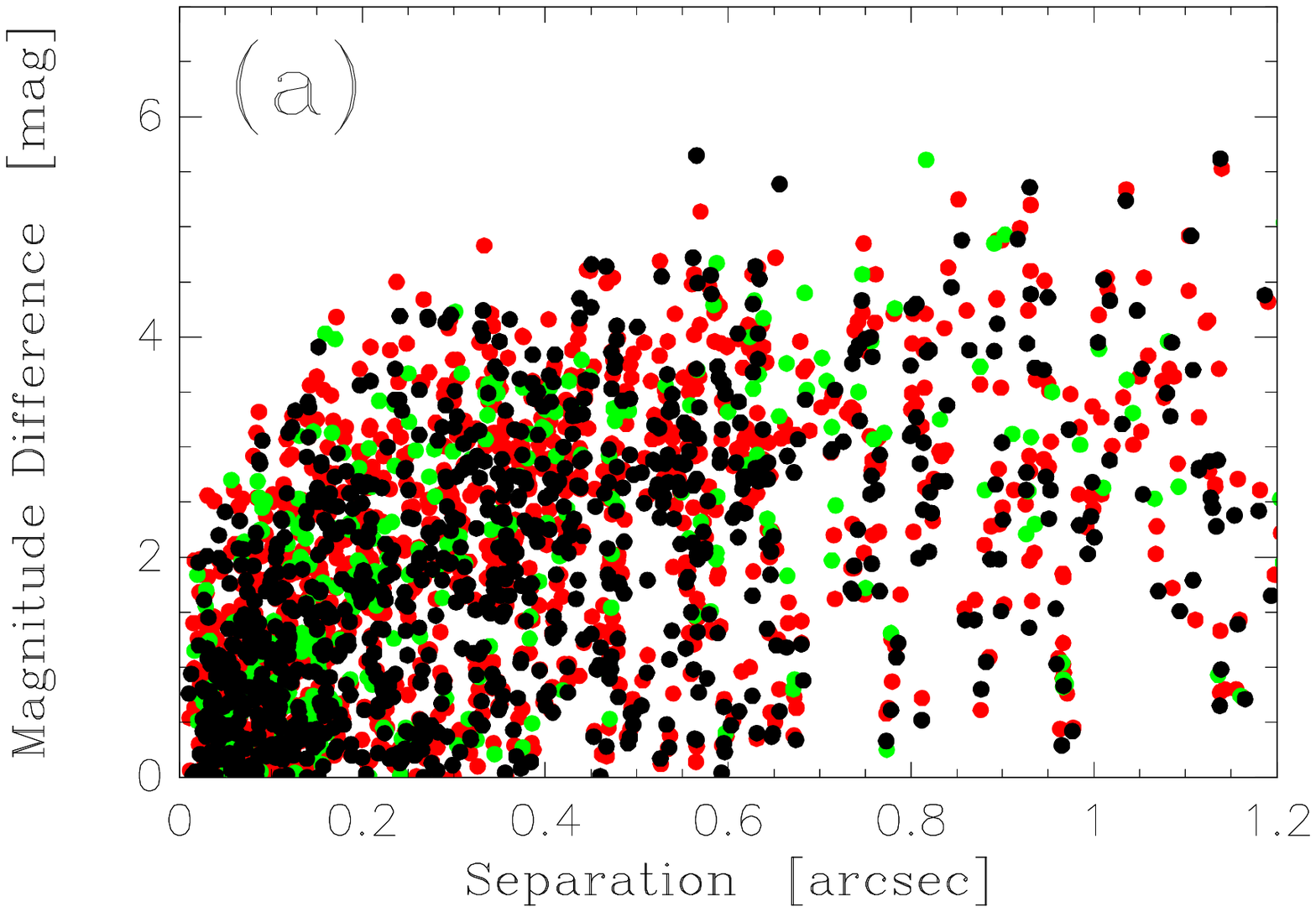}{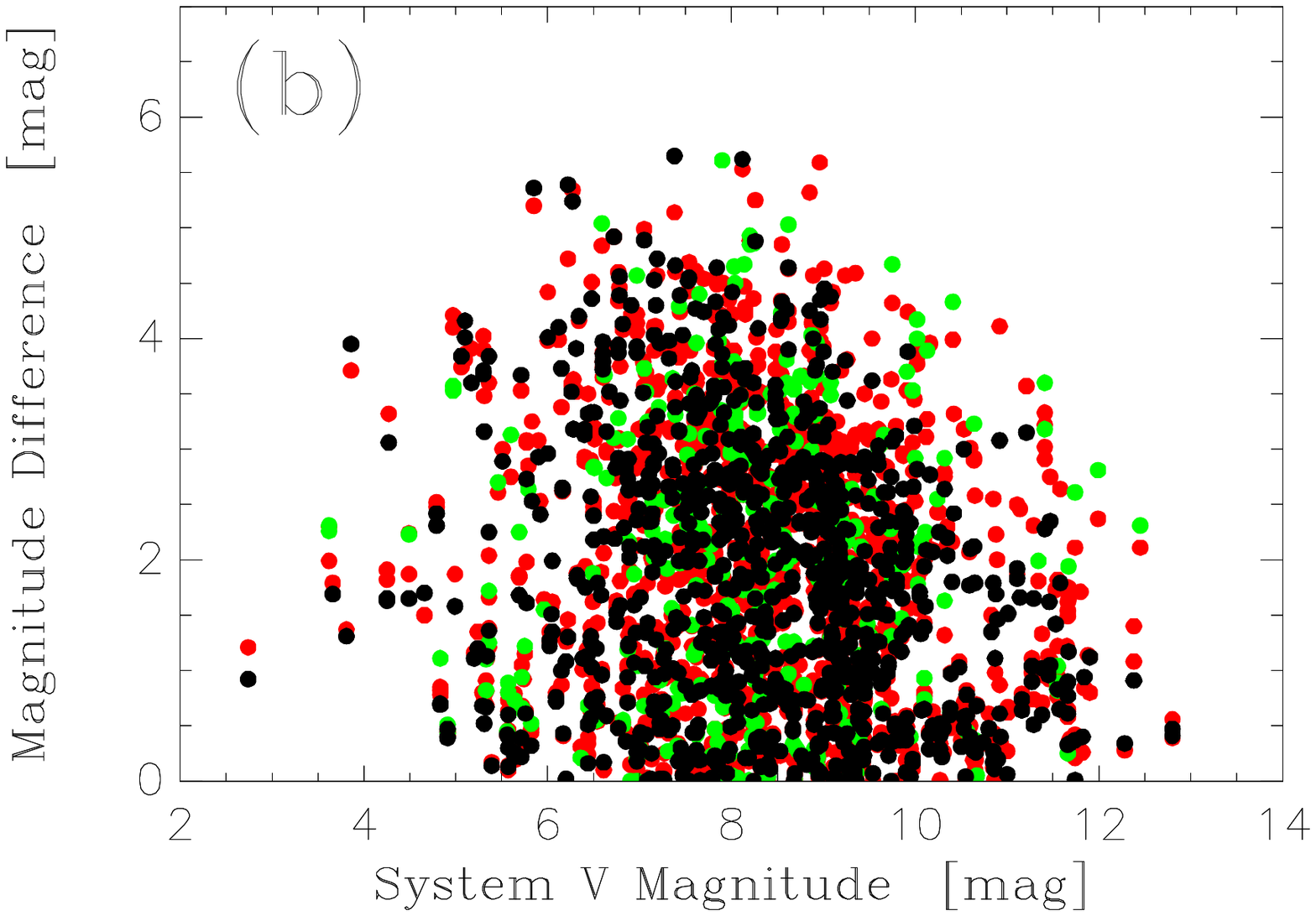}
\caption{
Magnitude difference as a function of (a) separation and (b) system
$V$ magnitude for the measures listed in Table 2. 
In both plots, the color of the data point indicates the wavelength of
observation, {\it i.e.\ }measures taken with the 562 nm filter are shown
in green, those taken with the 692 nm filter are shown in red, and those
taken in the 880 nm filter are shown as black.
}
\end{figure}

\clearpage
                                                                                
\begin{figure}[tb]
\plottwo{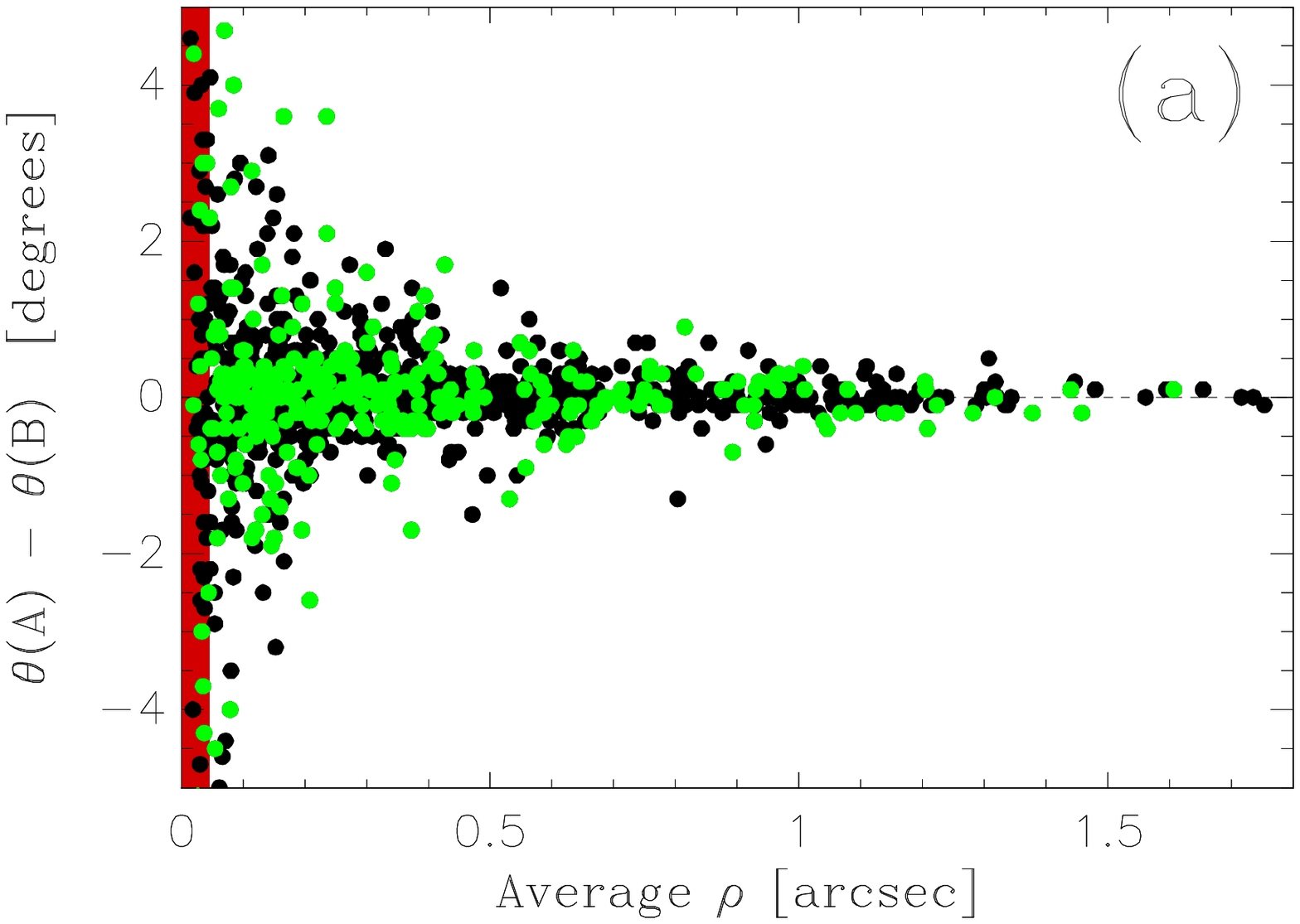}{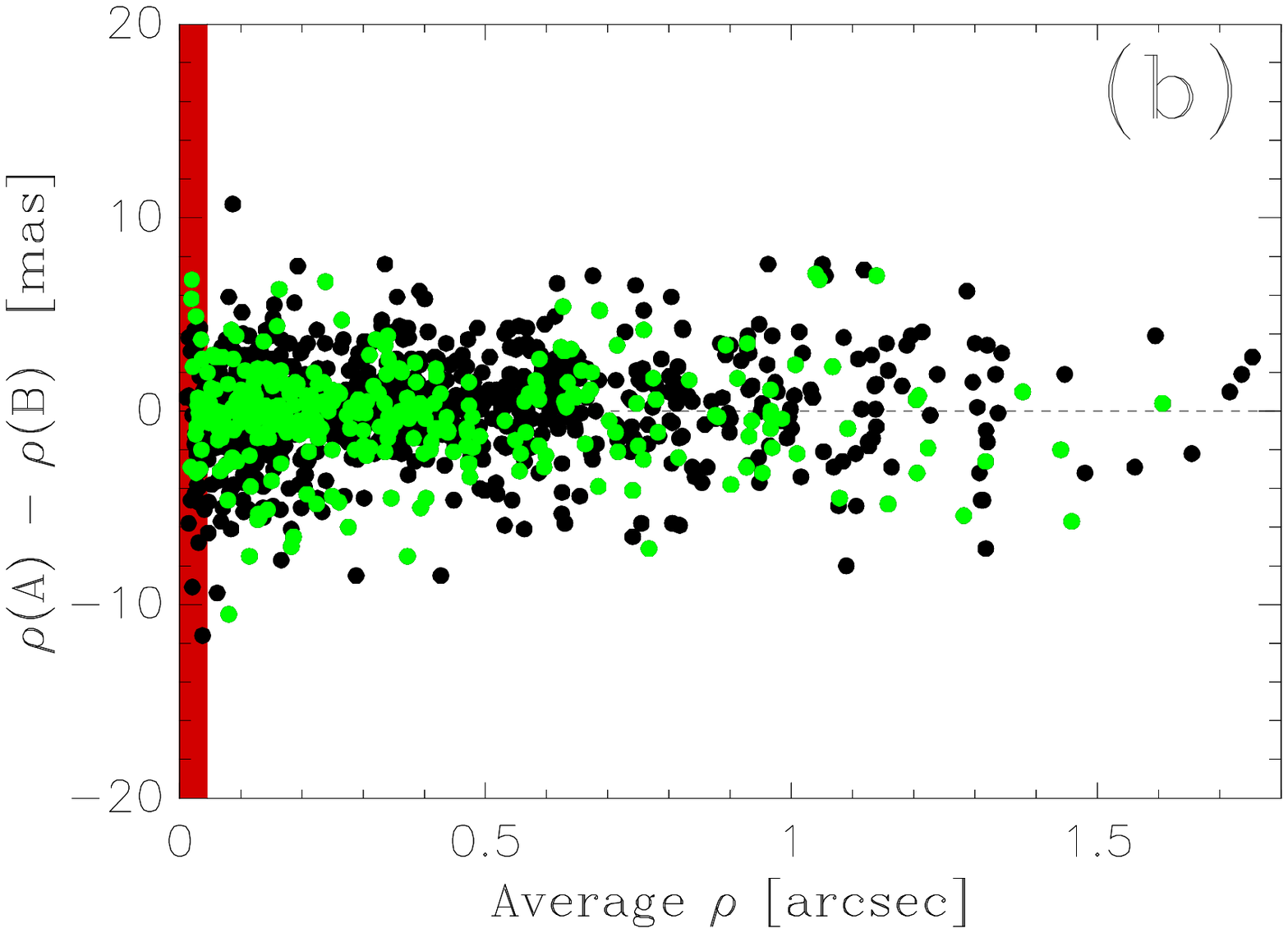}
\caption{
Measurement differences between the two channels of the instrument
plotted as a function of measured separation, $\rho$.
(a) Position angle ($\theta$) differences as a function of average separation.
(b) Separation ($\rho$) differences as a function of average separation.
In both plots, the red shaded area indicates the region below the
diffraction limit, and the color of the point indicates the filter used
in Channel B of the instrument, {\it i.e.\ }562 nm data are plotted as green points, and
880 nm  data are plotted as black points.
}
\end{figure}

\begin{figure}[tb]
\plottwo{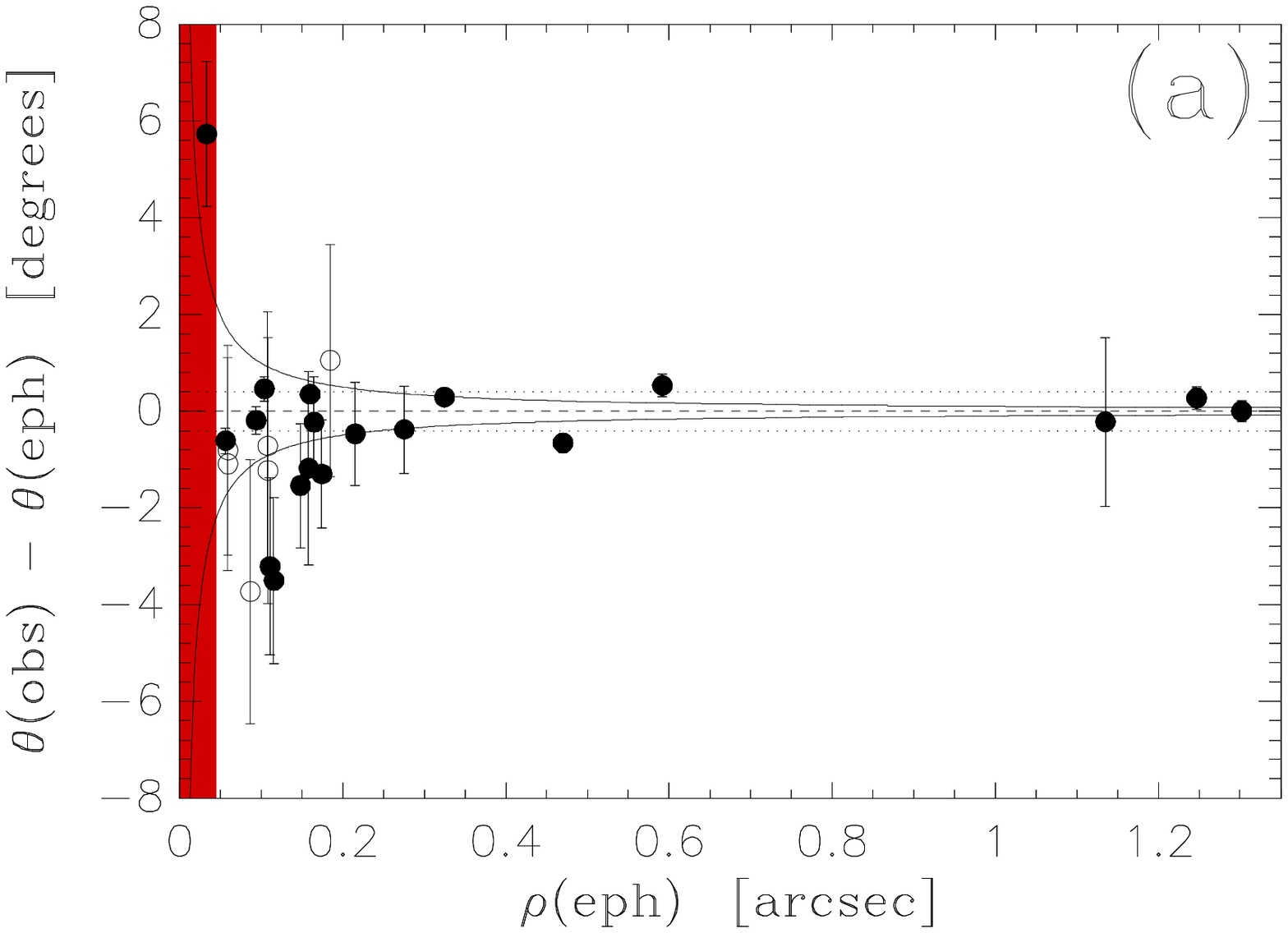}{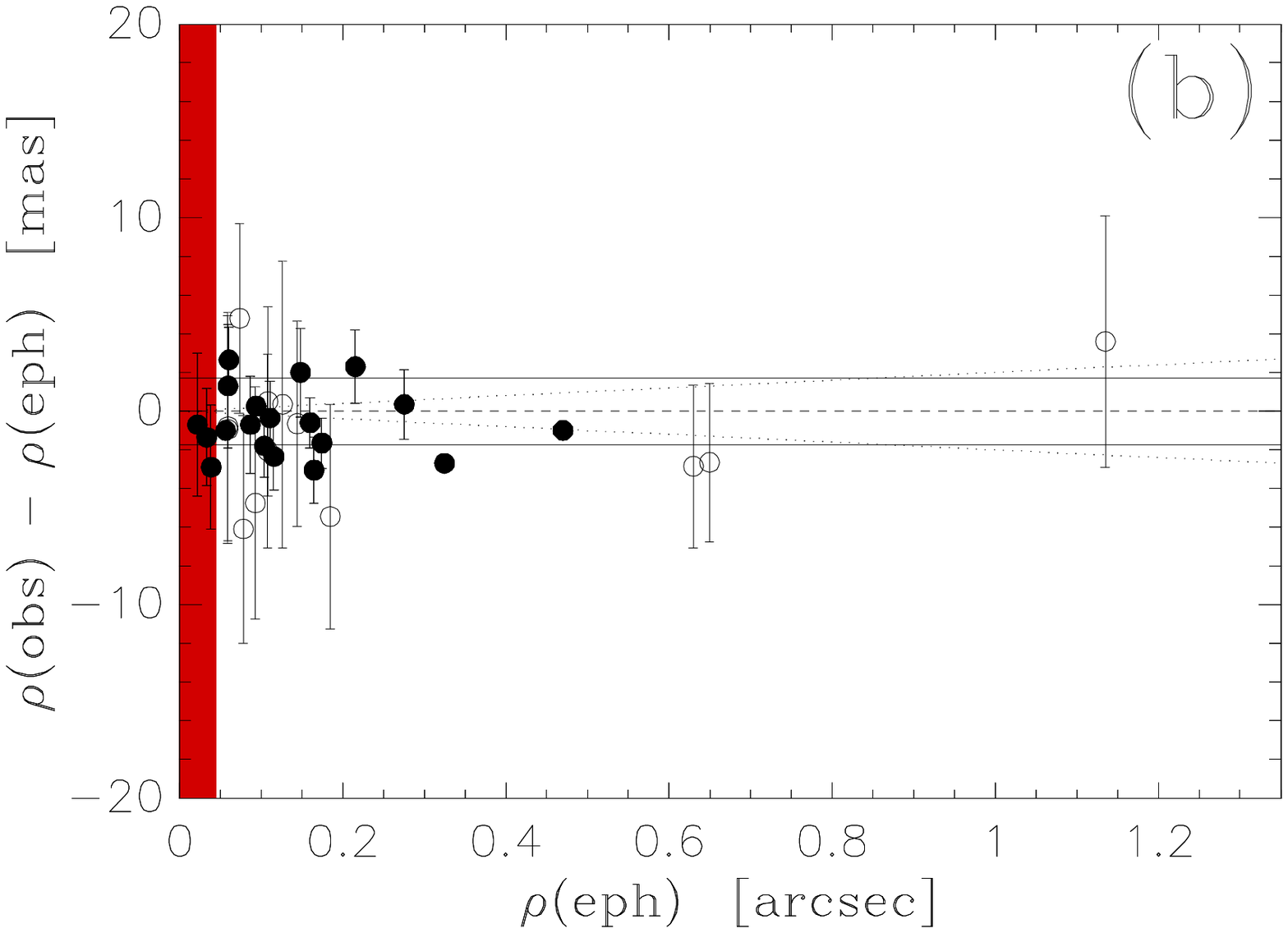}
\caption{
Residual plots for systems that have very well-determined 
orbits, listed in Table 3. 
(a) Position angle ($\theta$) residuals as a function of ephemeris separation
for systems with ephemeris uncertainties less than 4 degrees.
(b) Separation ($\rho$) residuals as a function of ephemeris separation
for systems with ephemeris uncertainties of less than 8 mas.
In both plots, the red shaded area indicates the region below the
diffraction limit, the dashed line is the zero line to guide the eye, 
the solid curves indicate the instrument repeatability as described
in the text, and the dotted curves are the uncertainty values from the 
scale calibration. For position angles, systems with uncertainties in the
orbital prediction of less than 2 degrees are plotted as filled.
For separations, systems with ephemeris uncertainties for the epoch of
observation of less than 4 mas are shown as filled.
}
\end{figure}

\clearpage

\begin{figure}
\plottwo{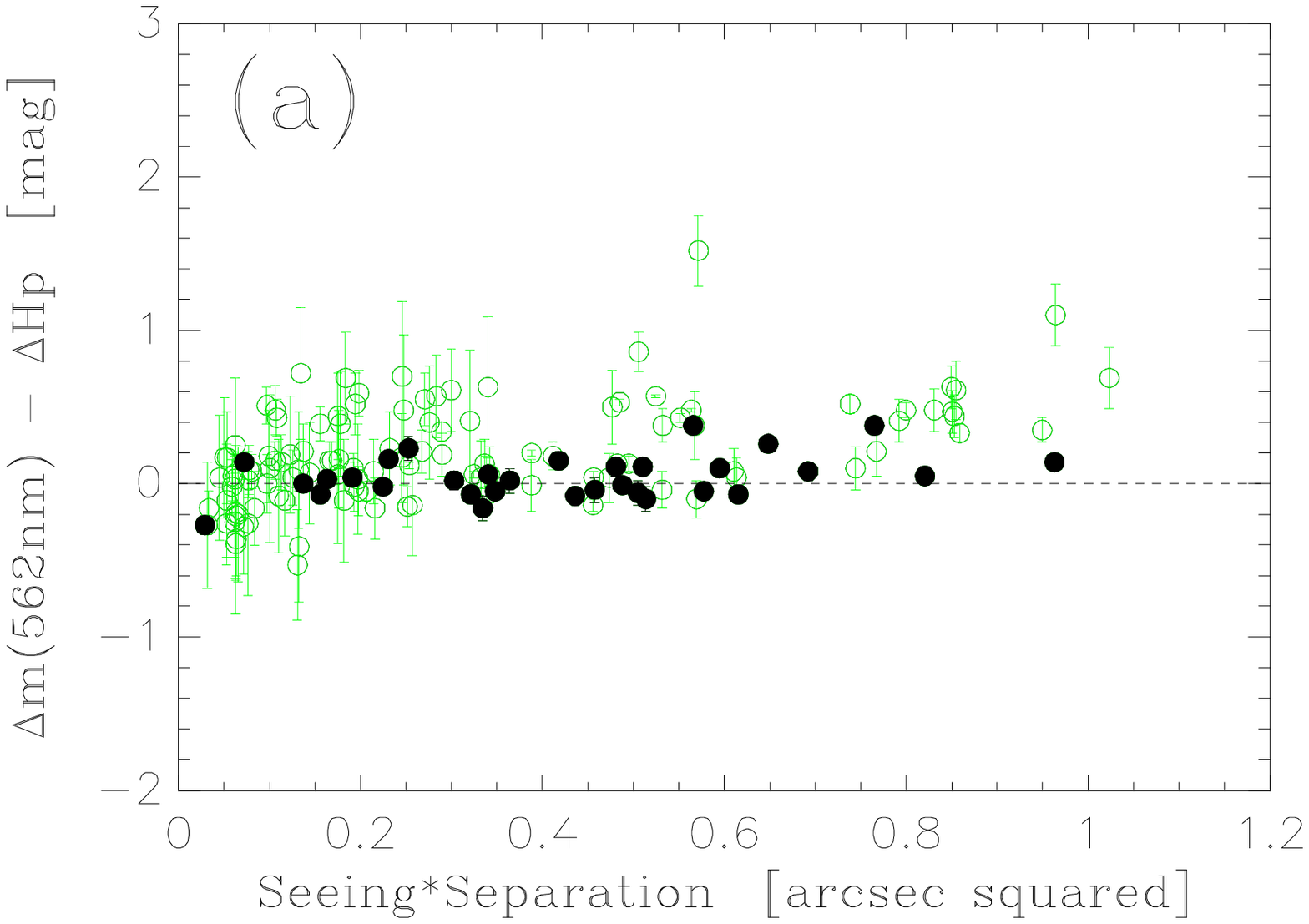}{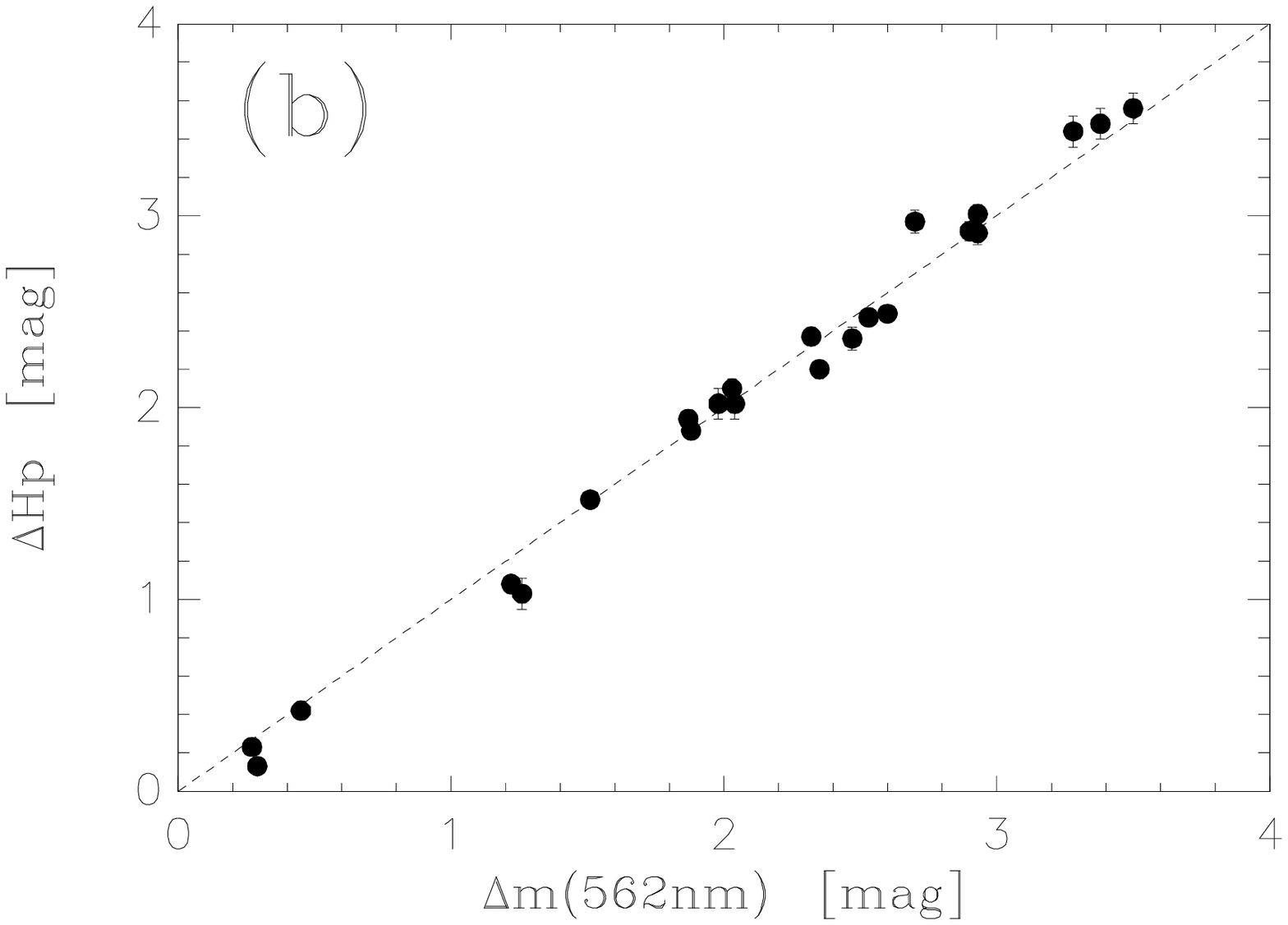}
\end{figure}
\begin{figure}
\plottwo{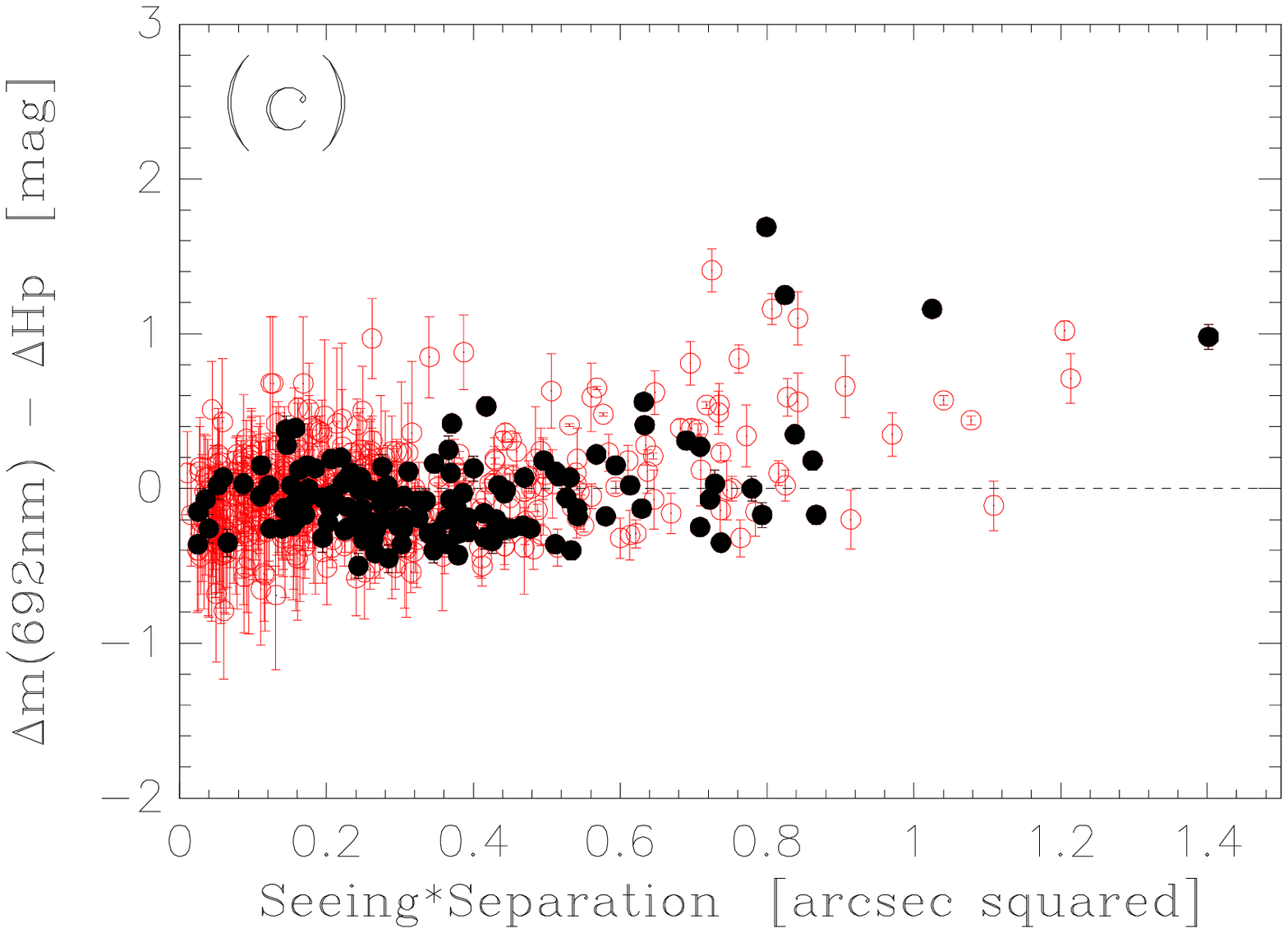}{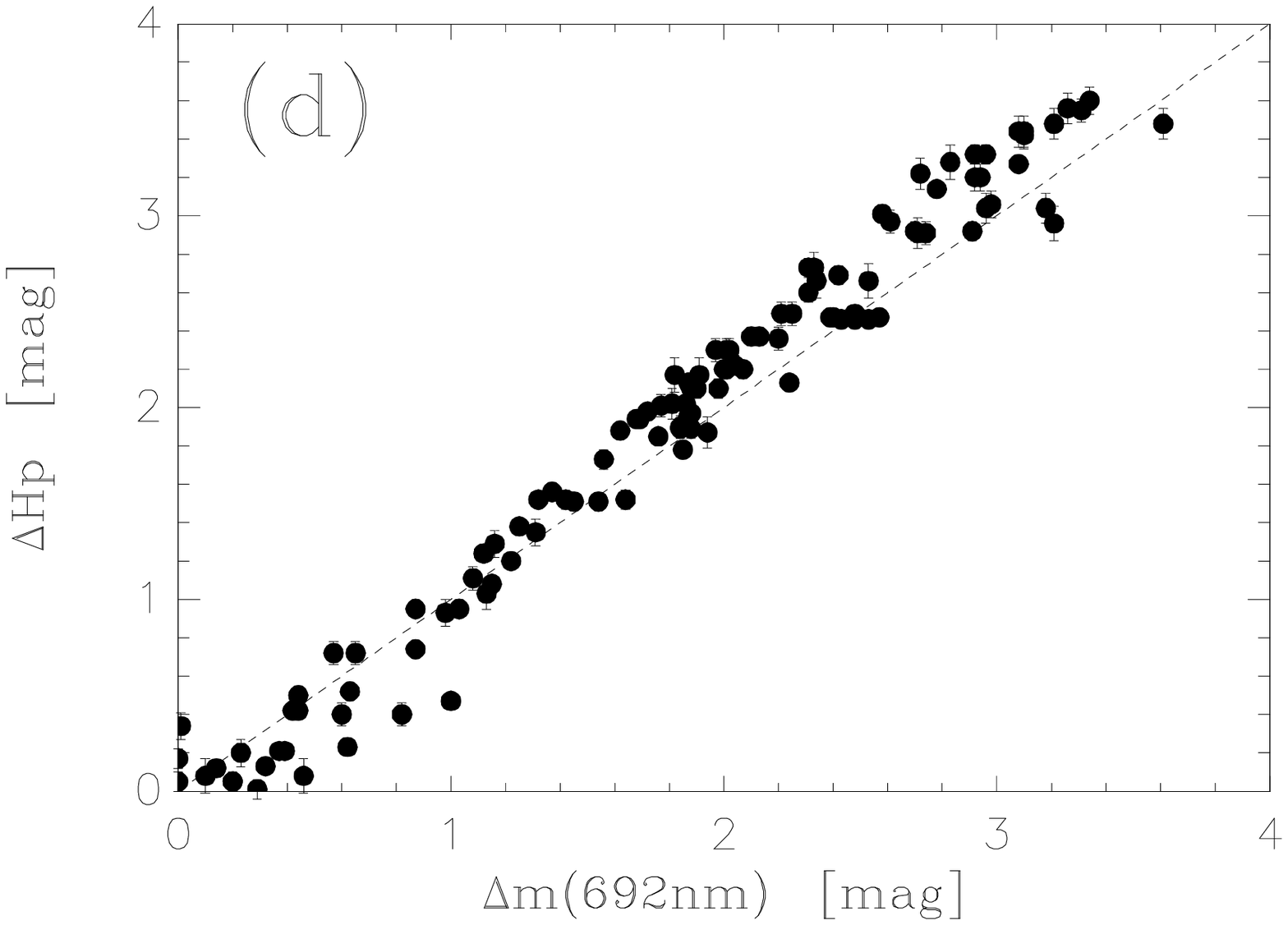}
\end{figure}
\begin{figure}
\plottwo{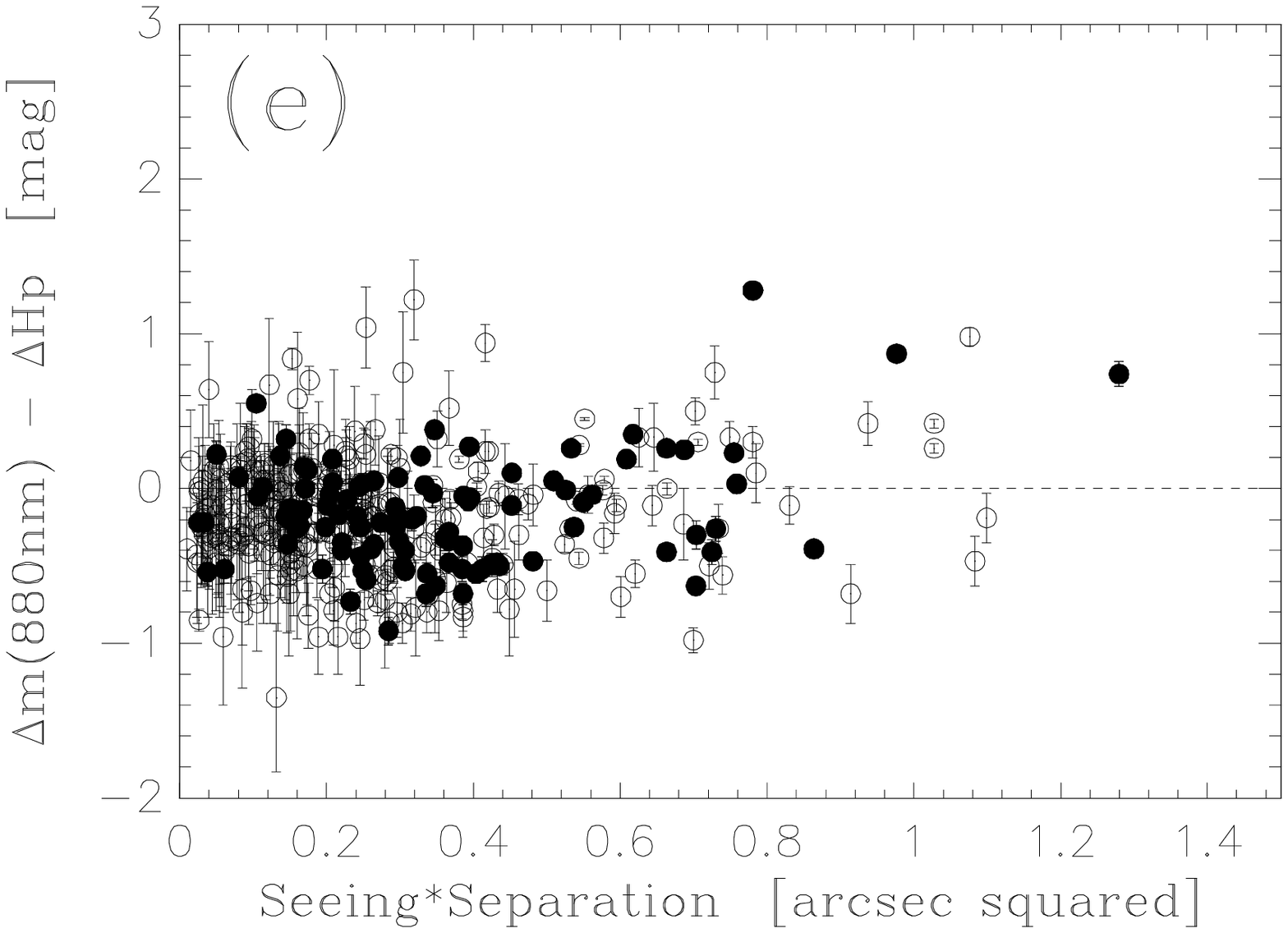}{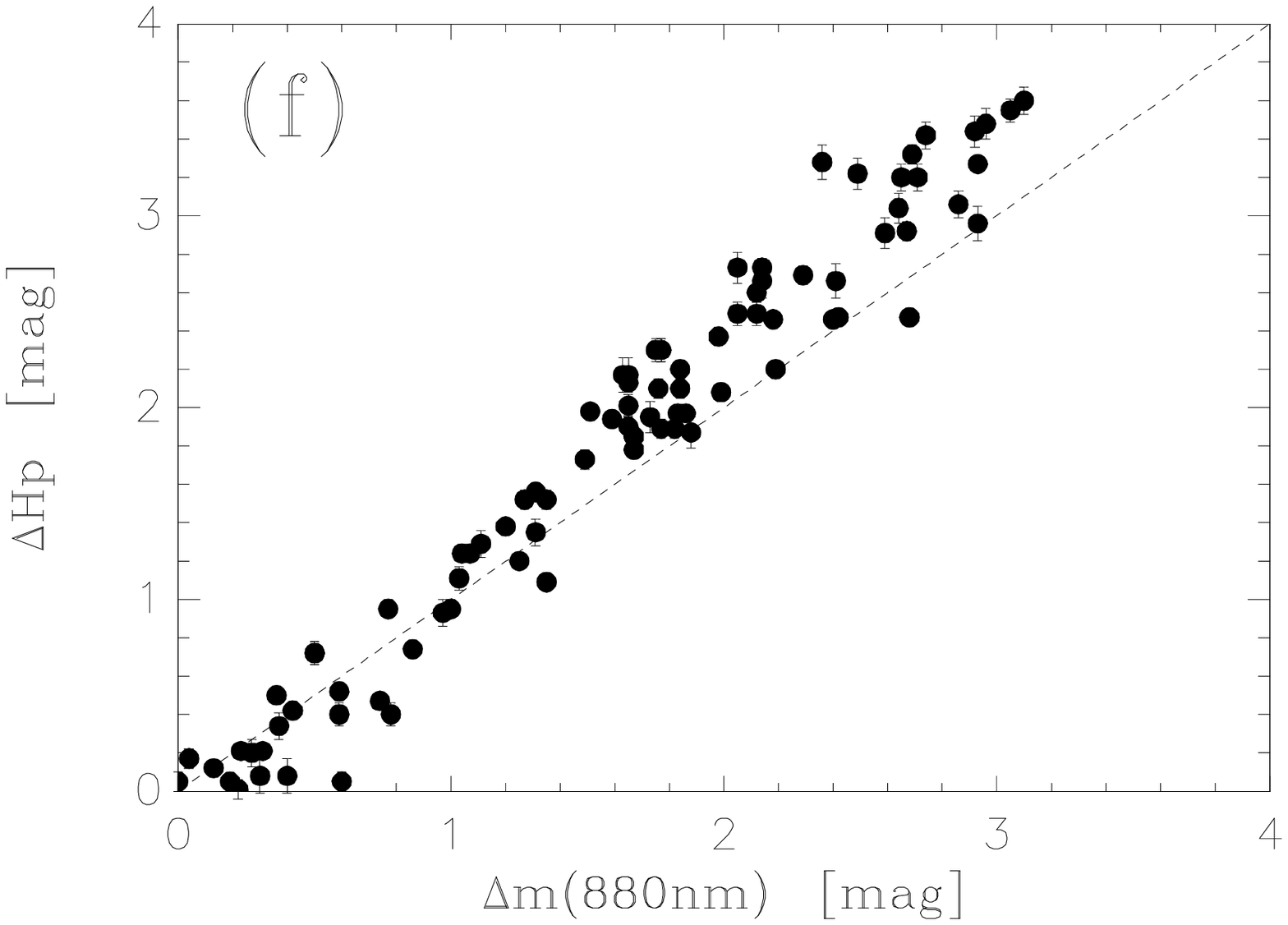}
\caption{
Photometric precision for the data set as a whole. In all plots, {\it Hipparcos}
measures with uncertainties less than 0.1 magnitudes are shown as filled 
black circles, and measures with larger uncertainties are
shown as colored open circles, with the color indicating the wavelength used.
(a) Differences in $\Delta m$ as a function of seeing times separation
for the 562-nm filter versus $\Delta H_{p}$.
(b) A plot of $\Delta H_{p}$ as a function of $\Delta m$ obtained at
562 nm.
(c) The same as panel (a) 
for the 692-nm filter.
(d) The same as panel (b) for the
692-nm filter.
(e) The same as panel (a)
for the 880-nm filter.
(f) The same as panel (b) for the
880-nm filter. 
}
\end{figure}

\begin{figure}
\plottwo{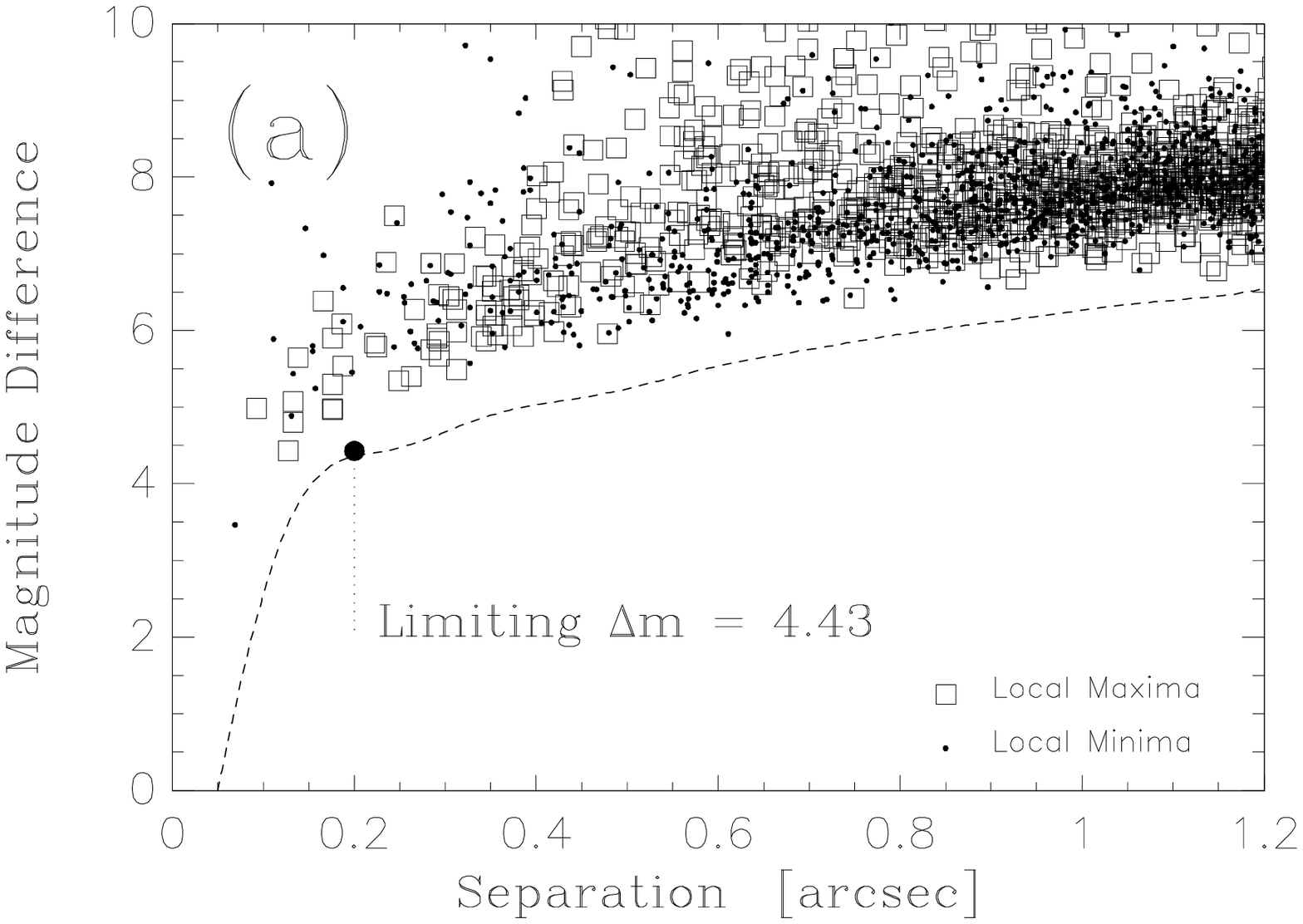}{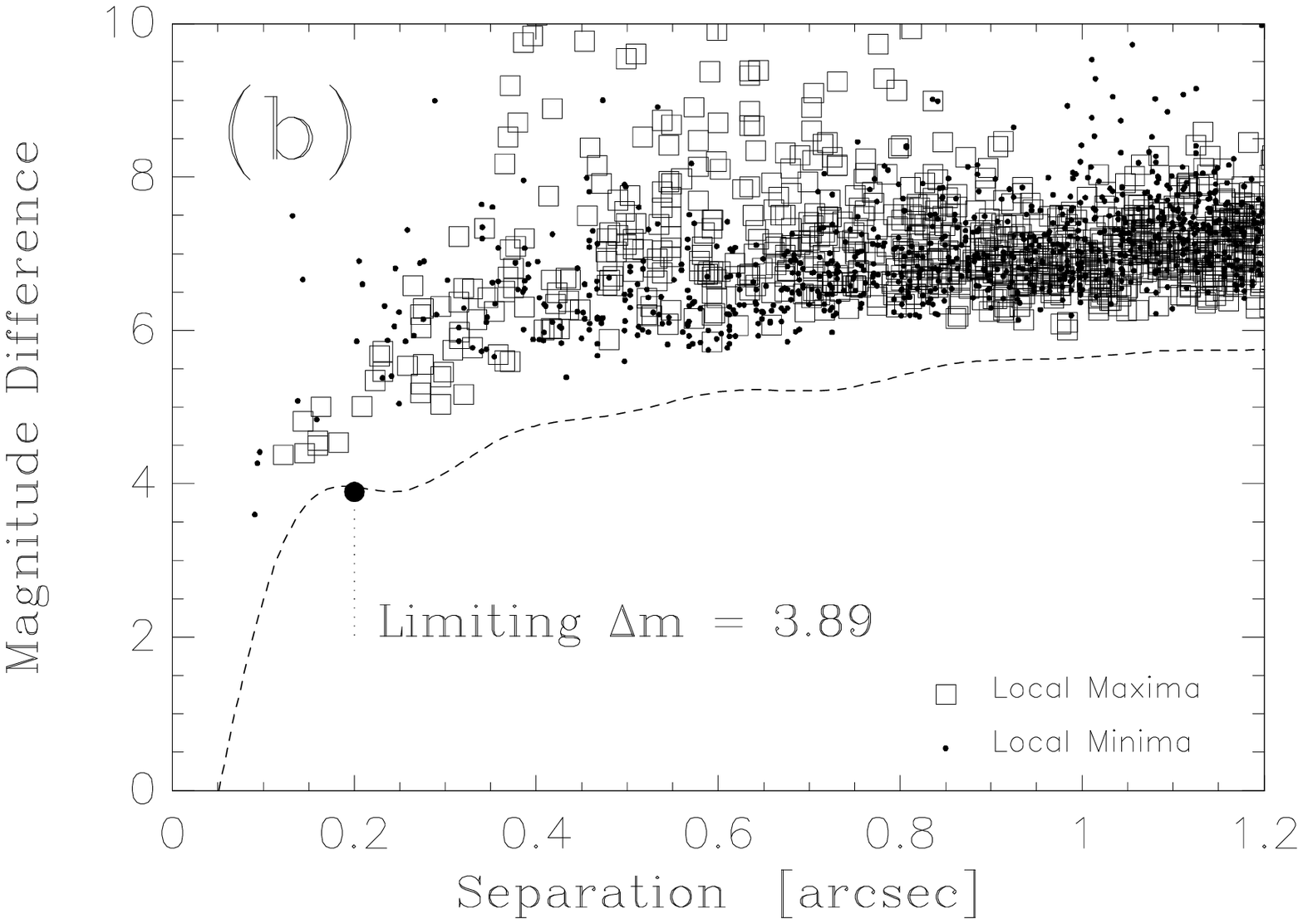}
\end{figure}
\begin{figure}
\plottwo{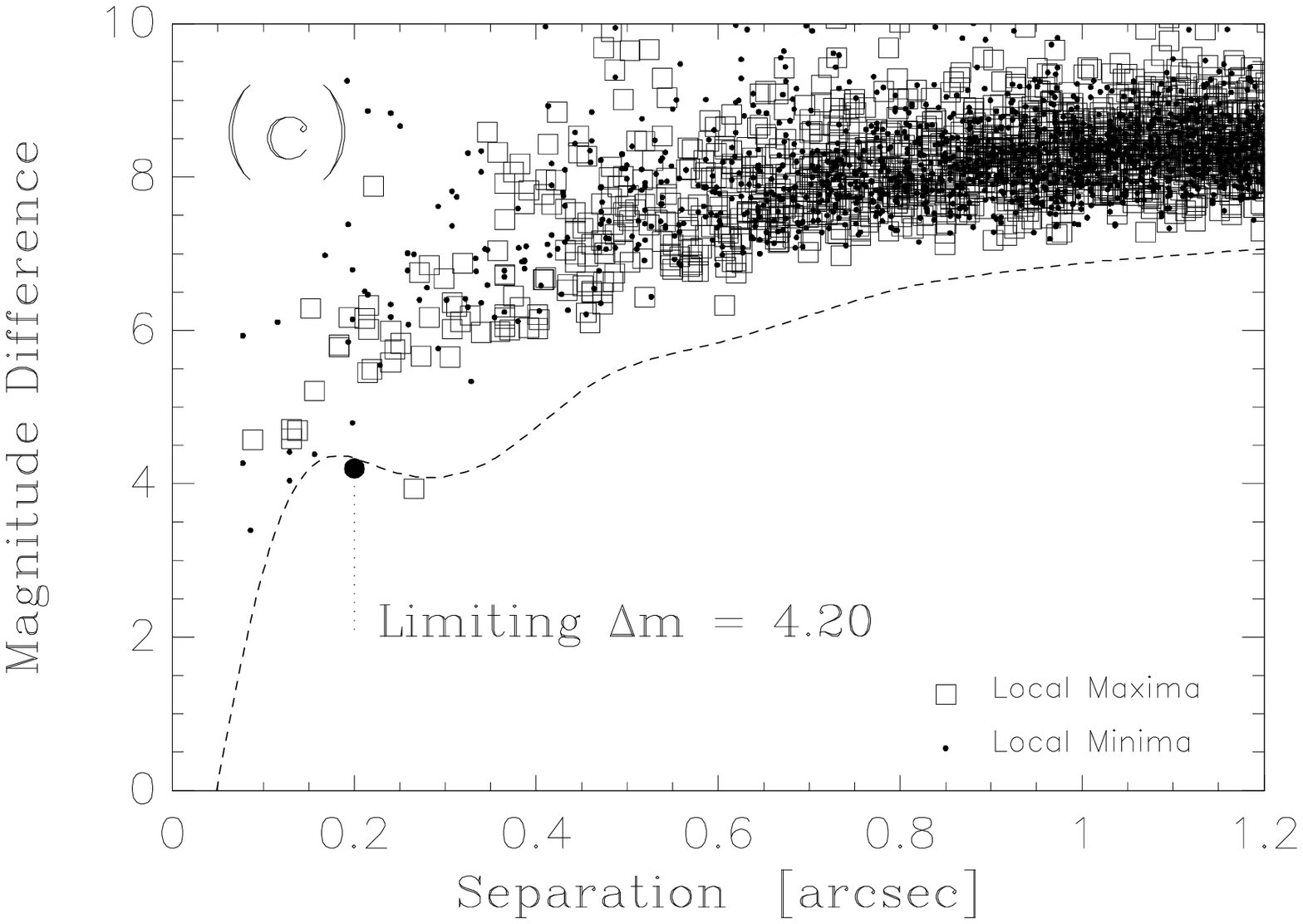}{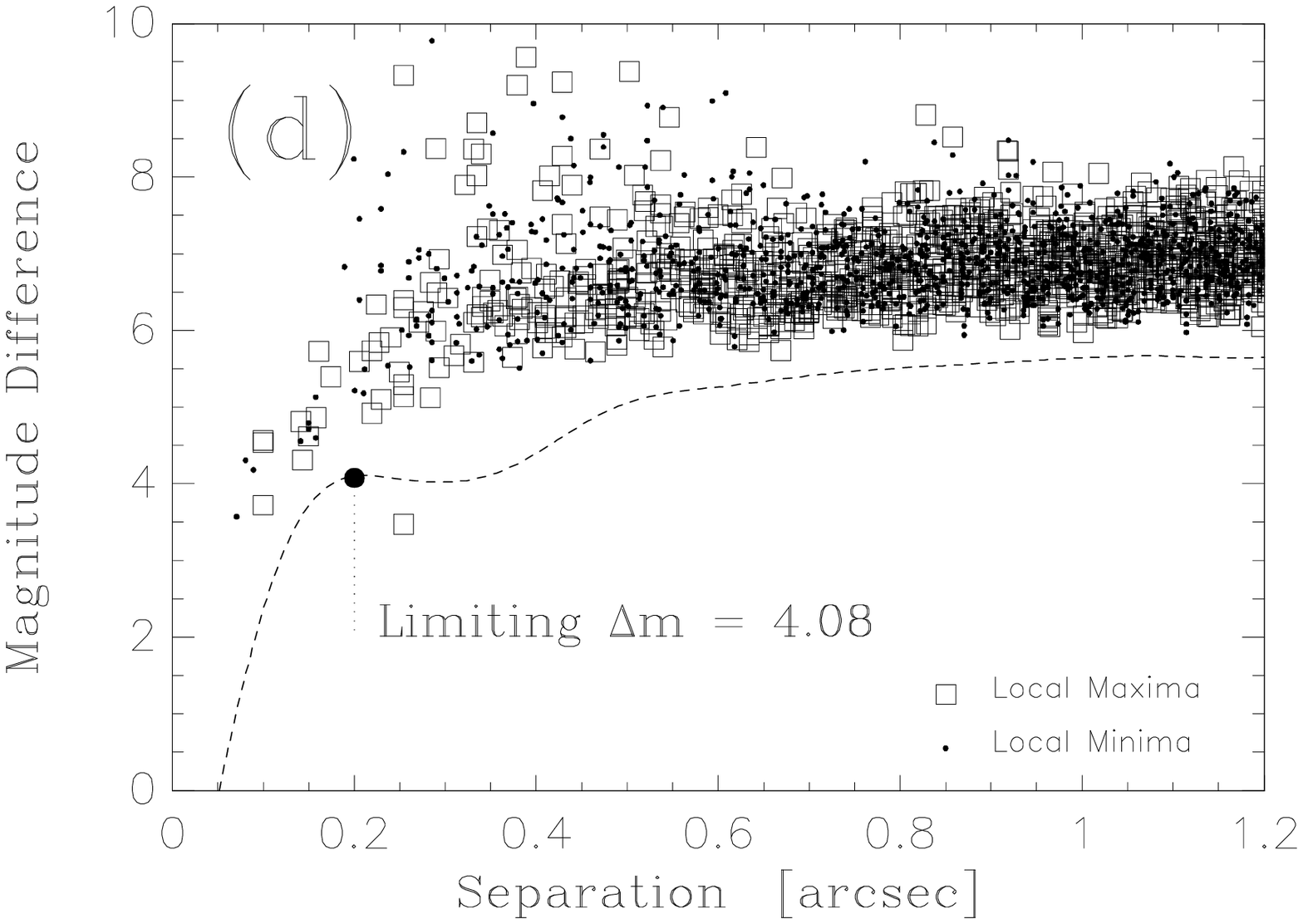}
\caption{
Detection limit plots for HIP 16467, where no companion was found, 
and HIP 31703, where we report a new component in Table 2.
In all cases squares represent the positions of local maxima in the 
reconstructed image and dots represent local minima (where the absolute
value of the minimum is used). The dashed line represents the 5-$\sigma$
line as a function of separation.
(a) HIP 16467, at 692 nm. (b) HIP 16467 at 880 nm. Note that in this
case no point lies below the 5-$\sigma$ curve, indicating that no 
companion was found.
(c) YSC 191 = HIP 31703, at 692 nm.
(d) YSC 191 = HIP 31703, at 880 nm. In this case, a 
single square lies below the 5-$\sigma$ line in both filters, 
indicating a detection of a companion
at a separation of aproximately 0.26 arc seconds.
}
\end{figure}

\begin{figure}
\plotone{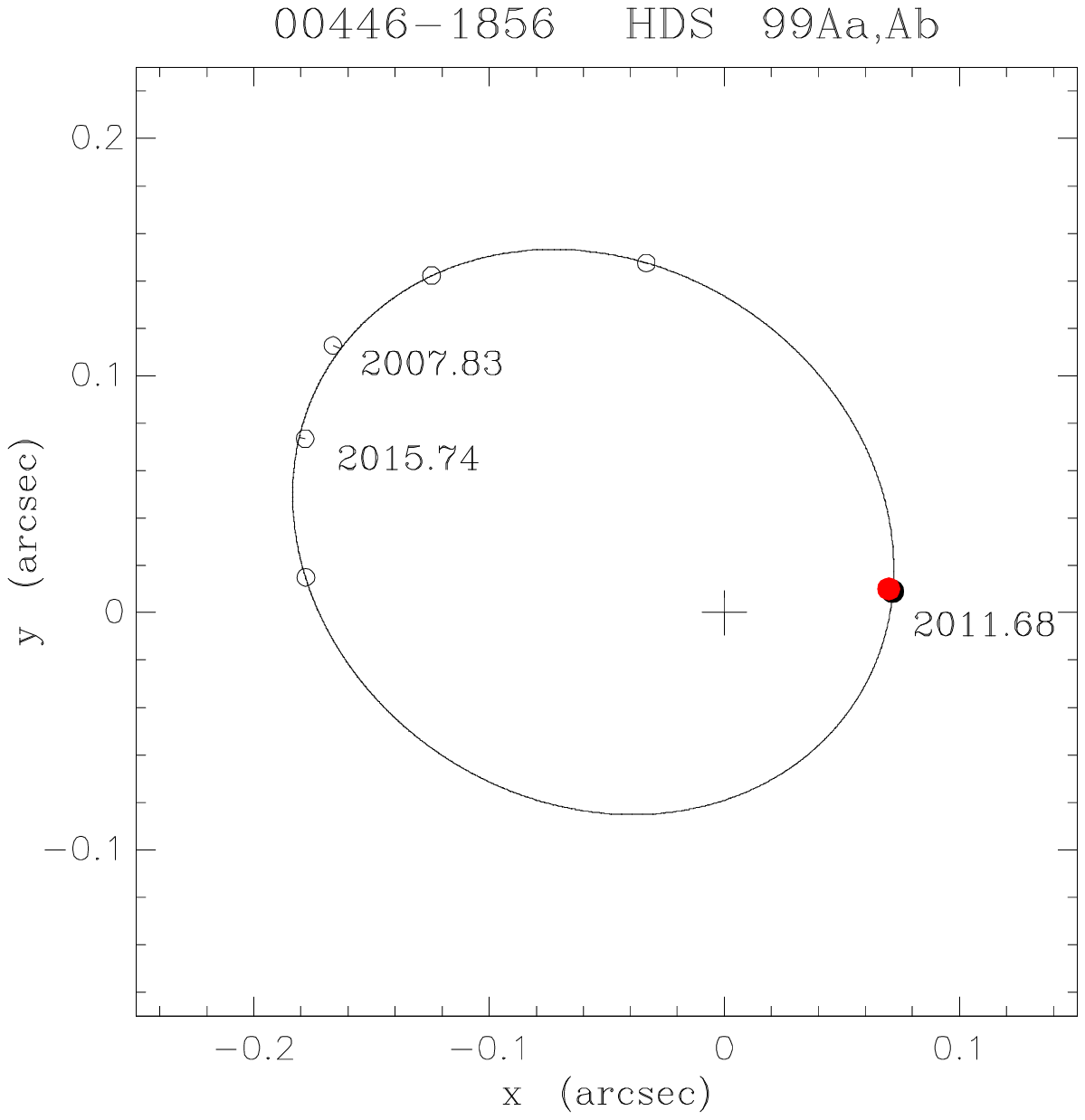}
\caption{
Astrometric data of HDS 99Aa,Ab together with the orbit presented in Table
5. The first and last epoch used in the orbit are labeled, as are the points
appearing in Table 2. Points shown as open circles are from data in the
4th Interferometric Catalog; points shown as filled circles indicate data
from Table 2, where the color of the point indicates the wavelength of the
observation (red being 692 nm and black being 880 nm). In all cases, a 
line segment is drawn from the ephemeris position to the center of the 
observational point. North is down and East is to the right.
}
\end{figure}

\begin{figure}
\plotone{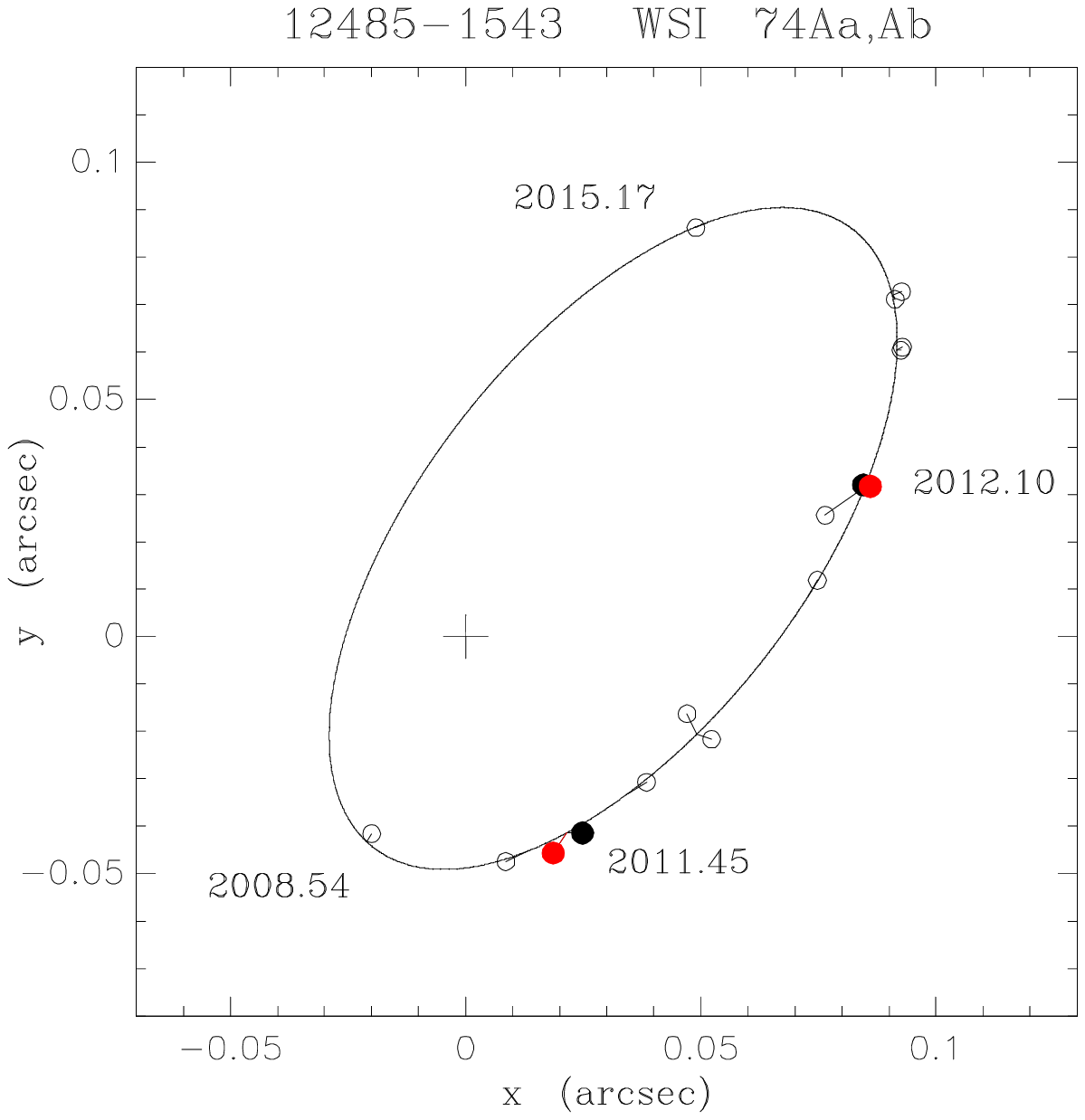}
\caption{
Astrometric data of WSI 74Aa,Ab together with the orbit presented in Table
5. The first and last epoch used in the orbit are labeled, as are the points
appearing in Table 2. Points shown as open circles are from data in the
4th Interferometric Catalog; points shown as filled circles indicate data
from Table 2, where the color of the point indicates the wavelength of the
observation (red being 692 nm and black being 880 nm). In all cases, a 
line segment is drawn from the ephemeris position to the center of the 
observational point. The quadrant of the 2011 observations shown in Table 2
is inconsistent with other existing measures; we have reversed that here and in
the orbit calculation. North is
down and East is to the right.
}
\end{figure}

\begin{figure}
\plotone{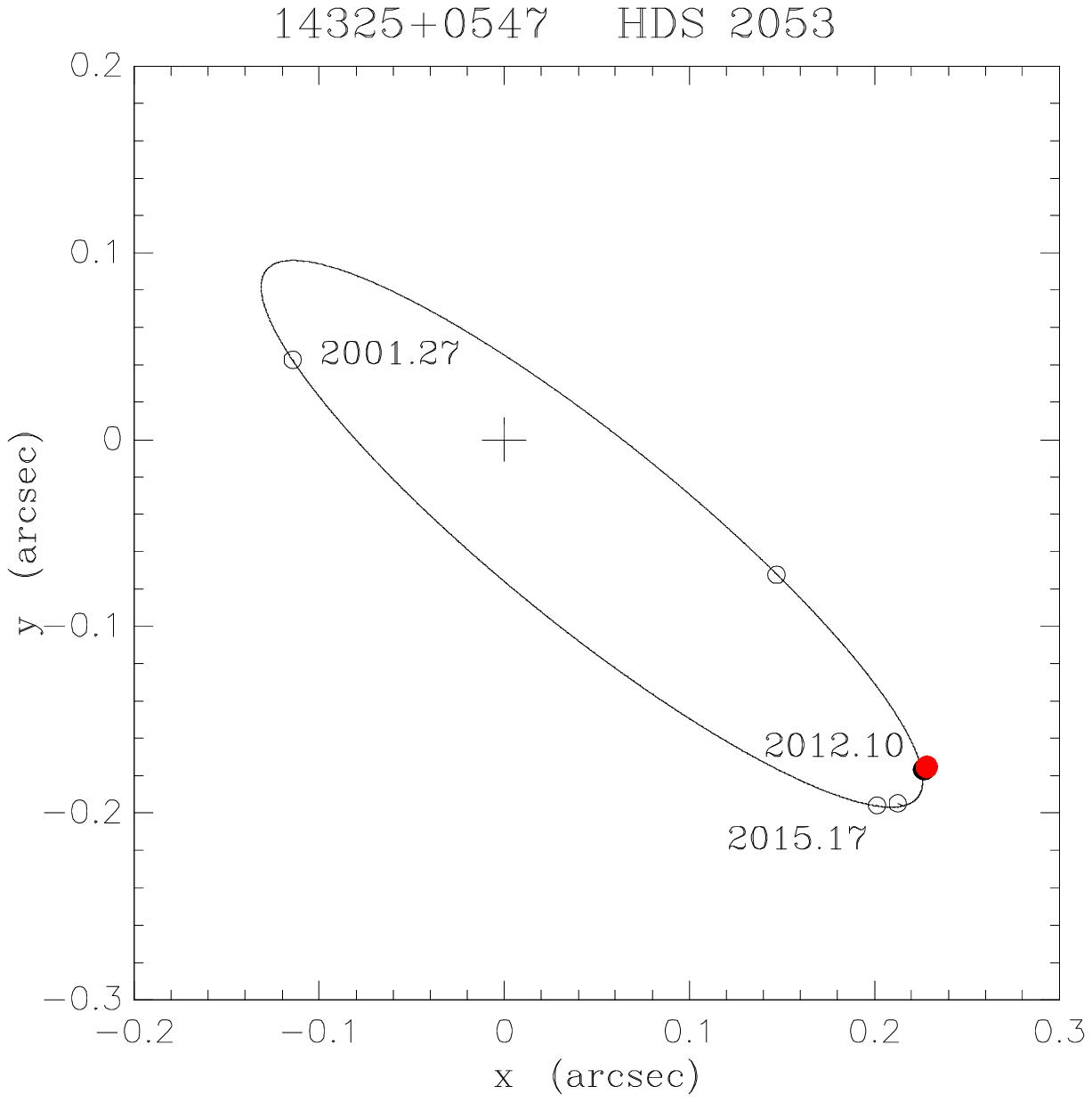}
\caption{
Astrometric data of HDS 2053 together with the orbit presented in Table
5. The first and last epoch used in the orbit are labeled, as are the points
appearing in Table 2. Points shown as open circles are from data in the
4th Interferometric Catalog; points shown as filled circles indicate data
from Table 2, where the color of the point indicates the wavelength of the
observation (red being 692 nm and black being 880 nm). In all cases, a 
line segment is drawn from the ephemeris position to the center of the 
observational point. North is
down and East is to the right.
}
\end{figure}

\end{document}